\documentclass[epj]{svjour}
\usepackage{latexsym}
\usepackage{graphics}
\usepackage[latin1]{inputenc}
\begin{document}
\renewcommand{\textfraction}{0.00000000001}
\renewcommand{\floatpagefraction}{1.0}
\title{Photoproduction of $\boldmath\eta\unboldmath$-mesons off nuclei for
$\boldmath E_{\gamma}\leq$ 2.2$\unboldmath$\ GeV}
\author{
        T.~Mertens\inst{1},
        I.~Jaegle\inst{1},
        P.~M\"uhlich\inst{2},
        J.C.S.~Bacelar\inst{3},
        B.~Bantes\inst{4},
        O.~Bartholomy\inst{5},
        D.E.~Bayadilov\inst{5,6},
        R.~Beck\inst{5},
        Y.A.~Beloglazov\inst{6},
        R.~Castelijns\inst{3},
        V.~Crede\inst{5,7},
        H.~Dutz\inst{4},
        A.~Ehmanns\inst{5},
        D.~Elsner\inst{4},
        K.~Essig\inst{5},
        R.~Ewald\inst{4},
        I.~Fabry\inst{5},
        K.~Fornet-Ponse\inst{4},
        M.~Fuchs\inst{5},
        C.~Funke\inst{5},
        R.~Gothe\inst{4,9},
        R.~Gregor\inst{8},
        A.B.~Gridnev\inst{6},
        E.~Gutz\inst{5},
        S.~H\"offgen\inst{4},
        P.~Hoffmeister\inst{5},
        I.~Horn\inst{5},
        J.~Junkersfeld\inst{5},
        H.~Kalinowsky\inst{5},
        S.~Kammer\inst{4},
        V.~Kleber\inst{4},
        Frank Klein\inst{4},
        Friedrich Klein\inst{4},
        E.~Klempt\inst{5},
        M.~Konrad\inst{4},
        M.~Kotulla\inst{1,8},
        B.~Krusche\inst{1},
        M.~Lang\inst{5},
        J.~Langheinrich\inst{4,9},
        H.~L\"ohner\inst{3},
        I.V.~Lopatin\inst{6},
        J.~Lotz\inst{5},
        S.~Lugert\inst{8},
        D.~Menze\inst{4},
        J.G.~Messchendorp\inst{3},
        V.~Metag\inst{8},
        C.~Morales\inst{4},
        U.~Mosel\inst{2},
        M.~Nanova\inst{8},
        D.V.~Novinski\inst{5,6},
        R.~Novotny\inst{8},
        M.~Ostrick\inst{4,10},
        L.M.~Pant\inst{8,11},
        H.~van~Pee\inst{5,8},
        M.~Pfeiffer\inst{8},
        A.K.~Radkov\inst{6},
        A.~Roy\inst{8,12},
        S.~Schadmand\inst{8,13},
        C.~Schmidt\inst{5},
        H.~Schmieden\inst{4},
        B.~Schoch\inst{4},
        S.V.~Shende\inst{3},
        V.~Sokhoyan\inst{5},
        A.~S{\"u}le\inst{4},
        V.V.~Sumachev\inst{6},
        T.~Szczepanek\inst{5},
        U.~Thoma\inst{5,8},
        D.~Trnka\inst{8},
        R.~Varma\inst{8,12},
        D.~Walther\inst{4},
        C.~Weinheimer\inst{5,14},
        \and C.~Wendel\inst{5}
\newline(The CBELSA/TAPS collaboration)
\mail{B. Krusche, Klingelberstrasse 82, CH-4056 Basel, Switzerland,
\email{Bernd.Krusche@unibas.ch}}
}
\institute{Department Physik, Universit\"at Basel, Switzerland
           \and Institut f\"ur Theoretische Physik I, Universit\"at Giessen,
             Germany
           \and KVI, University of Groningen, The Netherlands
           \and Physikalisches Institut der Universit\"at Bonn, Germany
           \and Helmholtz-Institut f\"ur Strahlen- und Kernphysik
                der Universit\"at Bonn, Germany
           \and Petersburg Nuclear Physics Institute, Gatchina, Russia
           \and Department of Physics, Florida State University, Tallahassee,
           USA
           \and II. Physikalisches Institut, Universit\"at Giessen, Germany
	   \and present address: University of South Carolina, USA
           \and present address: University of Mainz, Germany
           \and on leave from Nucl. Phys. Division, BARC, Mumbai, India
           \and on leave from Department of Physics, Indian Institute of Technology
	        Mumbai, India
	   \and present address: Institut f\"ur Kernphysik, Forschungszentrum J\"ulich, Germany
	   \and present address: University of M\"unster, Germany
}
\authorrunning{T. Mertens et al.}
\titlerunning{Photoproduction of $\eta$-mesons off nuclei}

\abstract{Photoproduction of $\eta$ mesons off $^{12}$C, $^{40}$Ca, $^{93}$Nb,
and $^{nat}$Pb nuclei has been measured with a tagged photon beam
with energies between 0.6 and 2.2 GeV. The experiment was
performed at the Bonn ELSA accelerator with the combined setup of the Crystal
Barrel and TAPS calorimeters. It aimed at the in-medium properties of the
S$_{11}$(1535) nucleon resonance and the study of the absorption properties
of nuclear matter for $\eta$ mesons. Careful consideration was given to
contributions from $\eta\pi$ final states and secondary production mechanisms
of $\eta$-mesons e.g. from inelastic $\pi N$ reactions of intermediate pions.
The analysis of the mass number scaling shows that the nuclear absorption
cross section $\sigma_{N\eta}$ for $\eta$ mesons is constant over a wide range
of the $\eta$ momentum. The comparison of the excitation functions to data off
the deuteron and to calculations in the framework of a BUU-model show no
unexplained in-medium modifications of the S$_{11}$(1535).
} 
\maketitle

\section{Introduction}
The study of possible in-medium modifications of the properties of hadrons
is a challenge for both theory and experiment. In contrast to any other
composite system, most of the mass of hadrons is generated by dynamical
effects from the interaction of the quarks. An important role is played
by the spontaneous breaking of chiral symmetry, the fundamental symmetry
of QCD. The symmetry breaking is reflected in a non-zero expectation
value of scalar $q\bar{q}$ pairs in the vacuum, the chiral condensate.
However, model calculations (see e.g. Ref. \cite{Lutz_92}) indicate a
temperature and density dependence of the condensate which is connected to
a partial restoration of chiral symmetry. In this way, hadron in-medium
properties are closely connected to the non-perturbative aspects of low-energy
QCD. While a direct relation between the quark condensate and the in-medium 
masses and widths of hadrons is not known, an indirect relation connects
the QCD picture with the hadron picture by QCD sum rules.
In the hadron picture, the in-medium modifications arise from the coupling of mesons to
resonance - hole states and the coupling of the modified mesons to resonances.
The best known example is the treatment of the $\Delta$ in the framework of
the $\Delta$-hole model (see e.g. Ref. \cite{Oset_83,Koch_83}). The hadron
in-medium spectral functions for $\pi$-, $\eta$-, and $\rho$-mesons and baryon
resonances have been recently calculated by Post, Leupold, and Mosel
\cite{Post_04} in a self-consistent coupled channel approach.

The experimental investigation of hadron in-medium properties is
complicated by initial and/or final state interactions.
Since the present experiment uses photoproduction of mesons, no initial but
significant final state interaction effects must be considered.
Here, the investigation of these reactions also allows us to perform a
detailed study of the meson - nucleus interactions which are responsible for the
final state interaction \cite{Roebig_96,Krusche_04,Krusche_04a}.
In case of the short-lived $\eta$ meson the investigation of final state interaction
effects is almost the only possibility to study the $\eta$-nucleon interaction.

Experimentally, one of the clearest, although still not fully explained,
in-medium effects has been observed in the excitation function of the total
photoabsorption reaction \cite{Frommhold_92,Bianchi_93,Bianchi_94}.
The bump in the elementary cross sections around 700 MeV incident photon energy,
corresponding to the second resonance region, namely the excitation of the
P$_{11}$(1440), D$_{13}$(1520), and S$_{11}$(1535) resonances, is not seen
in the nuclear data.
Many different effects have been discussed in the literature including trivial
explanations like nuclear Fermi motion.
Fermi motion certainly contributes to the broadening of the structure but cannot
explain its complete disappearance.
Collisional broadening of the resonances due to additional decay channels like
$NN^{\star}\rightarrow NN$ has been studied in detail in the framework of transport
models of the Boltzmann-Uehling-Uhlenbeck (BUU) type (see e.g. \cite{Lehr_00})
and can also not fully explain the data.
The situation is complicated by the fact that already on the free nucleon
the second resonance bump consists of a superposition of reaction channels
with different energy dependencies \cite{Krusche_04}.
Inclusive reactions like total photoabsorption do not allow to study
in-medium properties of individual nucleon resonances.
A study of the partial reaction channels is desirable, but their experimental
identification is more involved, and final state interaction effects \cite{Krusche_04a}
as well as experimental bias due to the averaging over the nuclear density
\cite{Lehr_01} must be accounted for (see Ref. \cite{Krusche_05} for a recent
summary).
Of special interest are meson production reactions which are
dominated in the energy region of interest by one of the three resonances.
Single and double pion production reactions have been employed for the study
of the D$_{13}$ resonance in the nuclear medium \cite{Krusche_04,Krusche_01},
although up to now without conclusive results.

Photoproduction of $\eta$ mesons in the second resonance region is an
excellent tool for the study of the S$_{11}$(1535) resonance, which
completely dominates this reaction \cite{Krusche_97,Krusche_95}.
Photoproduction of $\eta$ mesons has been studied for the free proton in great 
detail
over a wide range of incident photon energies and for different observables
\cite{Krusche_95,Ajaka_98,Bock_98,Renard_02,Dugger_02,Crede_05,Nakabayashi_06,Bartholomy_07,Elsner_07}.
The quasi-free reaction off the neutron bound in light nuclei has been
investigated in detail for incident energies up to the peak position
of the S$_{11}$(1535) ($\approx$ 800 MeV) \cite{Krusche_95a,Hejny_99,Weiss_03},
quasi-free neutron/proton cross section ratios for a few angular ranges up to
photon energies of 1~GeV have been reported in \cite{Hoffmann_97} and the
coherent photoproduction off light nuclei has been investigated for the
deuteron and Helium isotopes \cite{Hejny_99,Hoffmann_97,Weiss_01,Pfeiffer_04}.
The combined result of these experiments (see \cite{Krusche_03} for a summary)
was, that up to photon energies of $\approx$900 MeV also on the neutron
the reaction is completely dominated by the S$_{11}$(1535) with a constant
cross section ratio $\sigma_n /\sigma_p\approx 2/3$. Only very recently,
results from the GRAAL, ELSA, and Tohoku experiments 
\cite{Kuznetsov_07,Krusche_07,Jaegle_08,Miyahara_07}
indicated a stronger contribution of a higher lying resonance to
$\gamma n\rightarrow \eta n$ than to $\gamma p\rightarrow \eta p$ for photon
energies above 1~GeV. 

A first search for possible in-medium effects on the
S$_{11}$ spectral function was done with the TAPS experiment at MAMI
\cite{Roebig_96}. However, the experiment covered only incident photon
energies up to 800 MeV, i.e. approximately up to the peak position of the
resonance. The experimental results were in good agreement with BUU-model
calculations (see e.g. \cite{Lehr_00}). Subsequently, measurements
at KEK \cite{Yorita_00} and Tohoku \cite{Kinoshita_06} extended the energy
range up to 1.1 GeV. The KEK experiment reported some collisional broadening
of the S$_{11}$ resonance. The Tohoku experiment pointed to a significant
contribution of a higher lying resonance to the $\gamma n\rightarrow n\eta$
reaction. However, none of these experiments covered the full line shape of
the S$_{11}$.

Here, we report the measurement of $\eta$ photoproduction off carbon, calcium,
niobium, and lead nuclei up to incident photon energies of 2.2 GeV, i.e.
throughout and beyond the S$_{11}$ resonance range. For comparison, the
reaction has been studied for the same energy range off deuterium which
provides an estimate for the average nucleon cross section.

\section{Experimental setup}
\label{sec:setup}
\begin{figure}[th]
\resizebox{0.48\textwidth}{!}{%
  \includegraphics{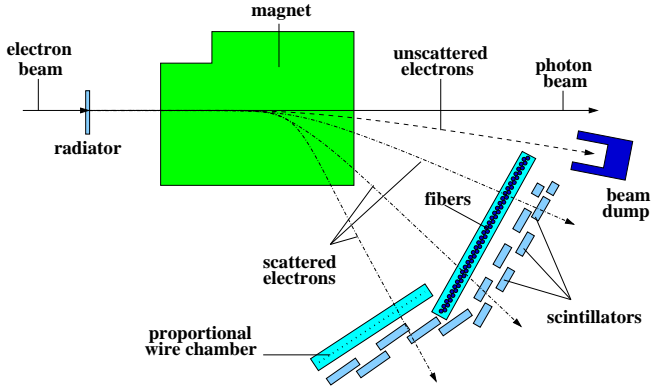}
}
\caption{Setup of the tagging spectrometer.
}
\label{fig:tagger}       
\end{figure}

The experiment was performed at the electron stretcher accelerator facility
ELSA \cite{Husmann_88,Hillert_06} in Bonn, using a 2.8~GeV electron beam. 
Real photons were produced by  Brems\-strahlung off a copper foil of  0.3 \% 
radiation length thickness. The photon energies were determined via the 
momentum analysis of the scattered electrons by a magnetic spectrometer. 
The tagging system, which is operated in coincidence with the production 
detector viewing the targets, is shown in Fig. \ref{fig:tagger}.
The direct electron beam is stopped in a beam dump while electrons having
emitted Bremsstrahlung are deflected into the detection system of the tagging facility.
The system
has 14 overlapping scintillator bars with 4 cm thickness which cover the
photon energy range between 22 \% to 95 \% of the incoming electron beam
energy $E_o$. Better energy resolution is provided by a scintillating fiber
detector which covers 18 \% to 80 \% of $E_o$ and a wire chamber (80 \% to 92
\%). In the present experiment only the scintillating fiber detector was
used which provides an energy bin width of $\approx$10 MeV for the lowest
incident photon energies around 650 MeV and 2 MeV at the high energy end of
2.2~GeV. The total rate in the tagging system was 8 - 10 MHz for an incident 
electron beam intensity of $\approx$ 1nA.

\begin{figure}[hbt]
\vspace*{0.cm}
\resizebox{0.24\textwidth}{!}{%
  \includegraphics{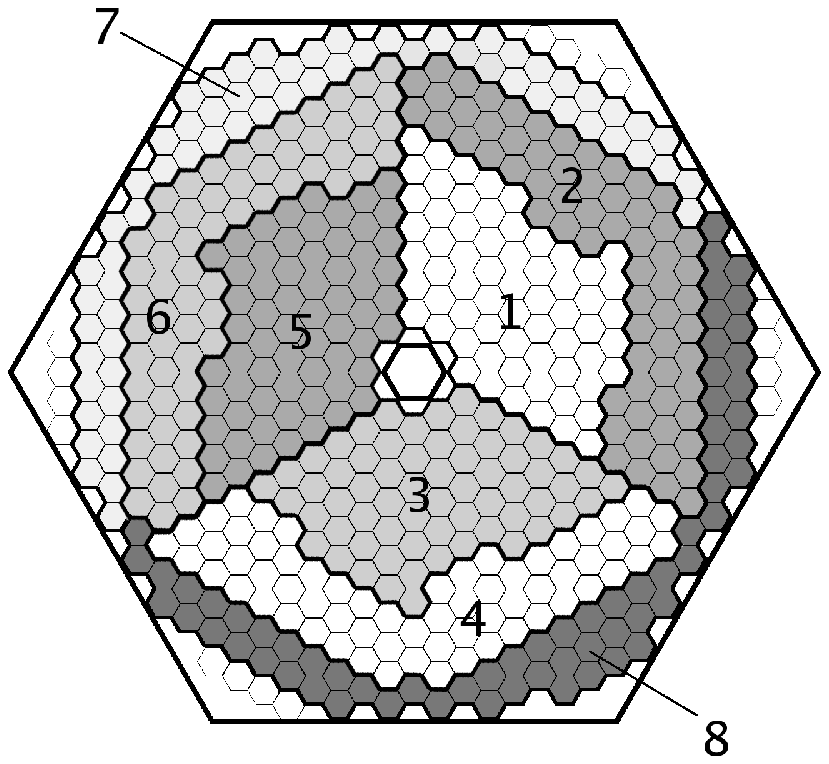}
}
\resizebox{0.24\textwidth}{!}{%
  \includegraphics{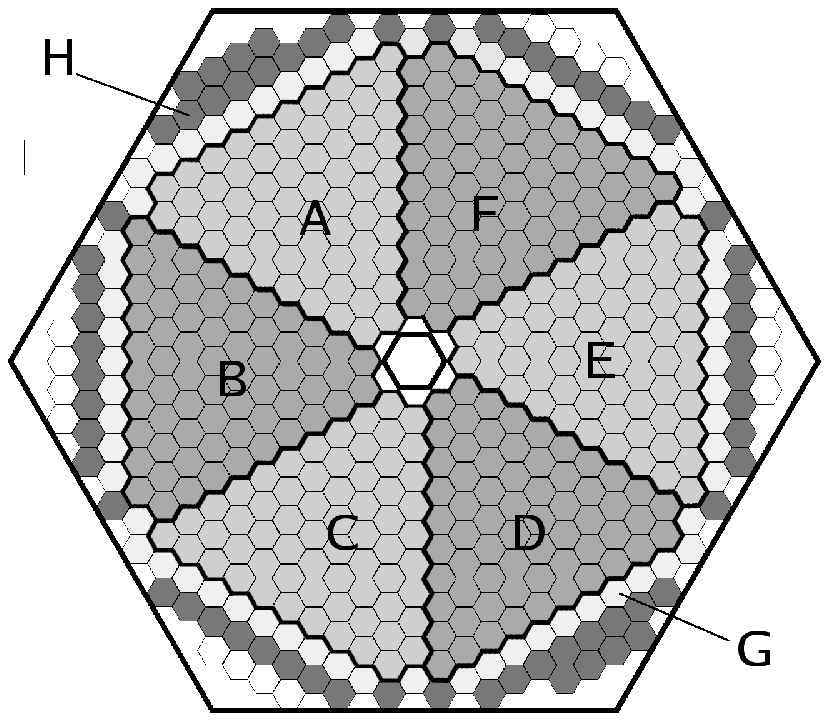}
}
\caption{Arrangement of the Crystal Barrel and TAPS detectors. Upper part:
side view, lower part: front view of the TAPS wall: left hand side: logical
segmentation for the LED-low trigger, right hand side: logical segmentation
for the LED-high trigger (see text).
}
\label{fig:calo}       
\end{figure}

Solid targets of $^{12}$C (20 mm length), $^{40}$Ca (10 mm), $^{93}$Nb
(1 mm), and $^{nat}$Pb (0.64 mm) were irradiated by the photon beam.
The lengths of the carbon, calcium, and lead targets corresponded to
8 - 10 \% of the respective radiation length $X_0$.
The niobium target was somewhat thicker ($\approx$ 17 \% of $X_0$),
all targets were 30 mm in diameter.
The $\eta$-mesons produced in the photonuclear reactions were
detected via their $\eta\rightarrow 3\pi^0\rightarrow 6\gamma$ decay
(branching ratio 32.5 \%) with a two-component electromagnetic calorimeter,
covering 99 \% of the full solid angle (see fig. \ref{fig:calo}).
The targets were mounted in the center of the Crystal Barrel detector
\cite{Aker_92} which covered the full azimuthal angle for polar angles
between 30$^{\circ}$ and 168$^{\circ}$. The Barrel consisted of 1290 CsI(Tl)
crystals of 16 radiation lengths $X_0$. Inside it, around the target, a
three-layer scintillating fiber detector \cite{Suft_05} (513 fibers of 2 mm
diameter, three layers oriented with respect to the $z$-axis by angles of
-24.5$^{\circ}$, +25.7$^{\circ}$, 0$^{\circ}$) was mounted for charged
particle identification. Compared to the standard setup of
the Crystal Barrel which was used for the measurement of $\eta$-photoproduction 
off the proton \cite{Crede_05} (see \cite{Pee_07} for a detailed description 
of the setup), the 90 forward-most crystals have been removed. The forward 
angular range down to 4.5$^{\circ}$ was covered by the TAPS detector
\cite{Novotny_91,Gabler_94}. This component consisted of 528 BaF$_2$ crystals
of hexagonal shape with an inner diameter of 5.9 cm and a length of 25 cm
corresponding to 12 radiation lengths. They were arranged in a wall-like
structure as shown in the lower part of fig. \ref{fig:calo}. A 5 mm thick
plastic scintillator was mounted in front of each BaF$_2$ crystal for the
identification of charged particles. The front face of the BaF$_2$ wall was
located 1.18 m from the center of the target. Both calorimeters have 
a comparable energy resolution of \cite{Aker_92,Gabler_94}
\begin{equation}
\frac{\sigma_E}{E}\approx \frac{2 - 3 \%}{\sqrt[4]{E/GeV}}\;\;\;.
\end{equation}
\begin{figure}[thb]
\resizebox{0.48\textwidth}{!}{%
  \includegraphics{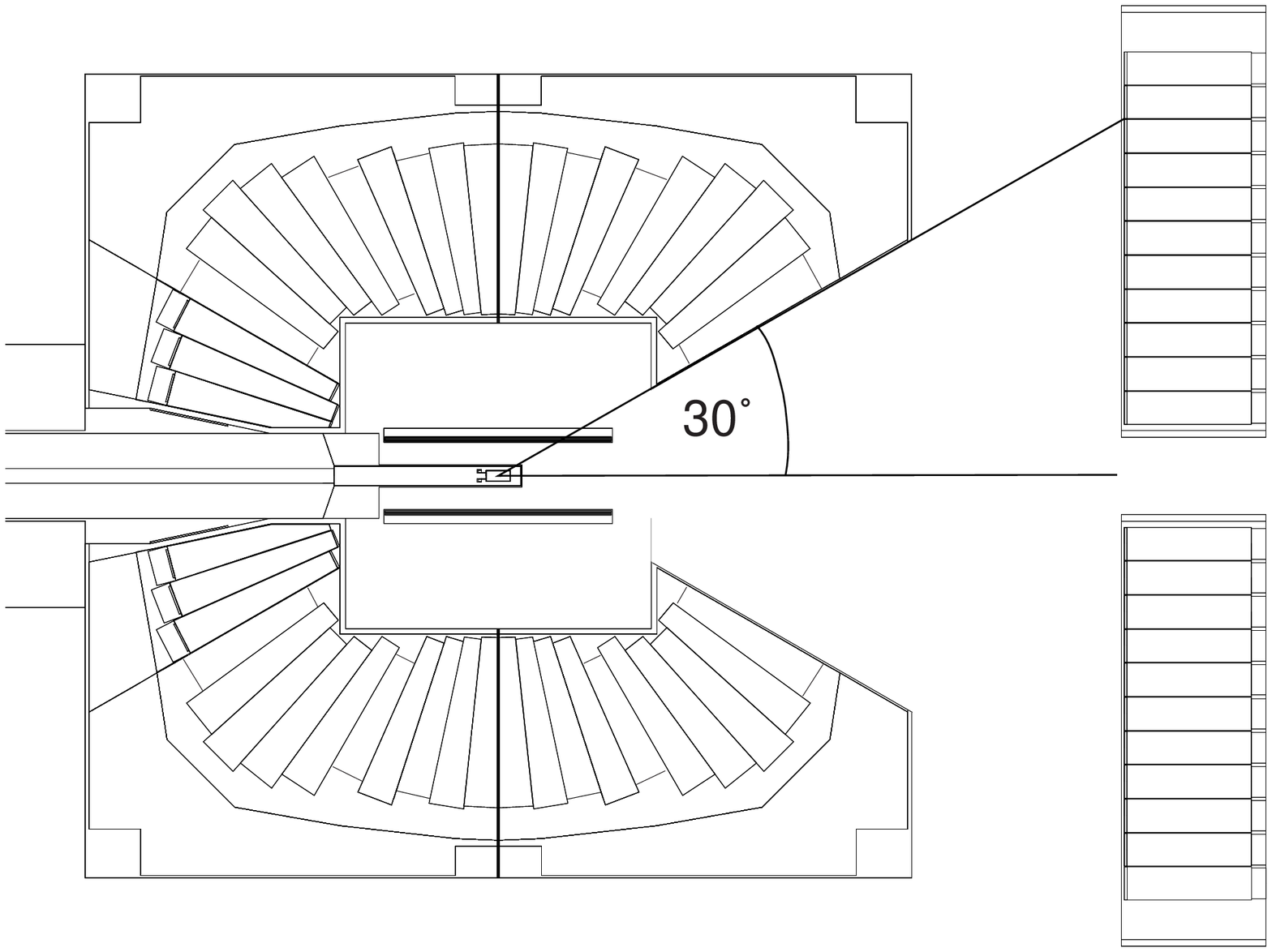}
}
\end{figure}
The impact points of photons are determined from the center of gravity of
the electromagnetic showers, so that the angular resolution is better than
the granularity of the crystals. It is 1.5$^{\circ}$ ($\sigma$) for the CB
\cite{Aker_92} for photons with energies above 50 MeV and 1.25$^{\circ}$
in TAPS. 
The fast BaF$_2$ modules were read out by photomultipliers, the CsI crystals 
by photodiodes. Therefore, only information from the TAPS wall could be used 
for the first level trigger. 
For this purpose each module of the TAPS wall was equipped with two independent 
leading edge discriminators which were combined in two different ways into 
logical groups (see Fig. \ref{fig:calo}). For the present experiment the 
thresholds of the first set of leading edge discriminators were set to 60 MeV 
(LED-low) and the thresholds of the second set to 80 MeV (LED-high). A valid 
first level trigger was accepted if either at least two logical groups of the 
low-threshold or at least one group of the high-threshold discriminators had 
fired. In the latter case, a second level trigger from the FAst Cluster Encoder 
(FACE) of the Crystal Barrel, indicating at least two separated hits in the 
Crystal Barrel, was required in addition.
Due to the trigger conditions only the decay channel into six photons
could be used for the detection of the $\eta$ mesons since the probability to
find both photons from a two photon decay in TAPS is almost negligible.
It should be noted that this restriction occurs only for measurements off nuclei
where the recoil nucleon can be a neutron.
In case of a proton target the recoil proton can provide the trigger.

\section{Data analysis}
\label{sec:datana}
In the experiment, $\eta$-mesons have been identified via the decay chain
$\eta\rightarrow\pi^0\pi^0\pi^0\rightarrow 6\gamma$. 
Events with six detected photons without a condition on further detected 
charged and/or neutral particles (recoil nucleons, pions) were selected. 
The photon reconstruction and identification in the Crystal Barrel is 
discussed in detail in Ref. \cite{Pee_07}.
It is based on a cluster search algorithm and uses the
information from the three layer scintillating fiber detector for rejection
of charged particles. The photon identification in TAPS is based on the
information from the charged particle veto detectors, on a time-of-flight 
analysis, and on a pulse-shape analysis of the BaF$_2$ signals. 
Details of this analysis procedure can be found in Ref. \cite{Bloch_07}.

\begin{figure}[ht]
\centerline{\resizebox{0.45\textwidth}{!}{%
\includegraphics{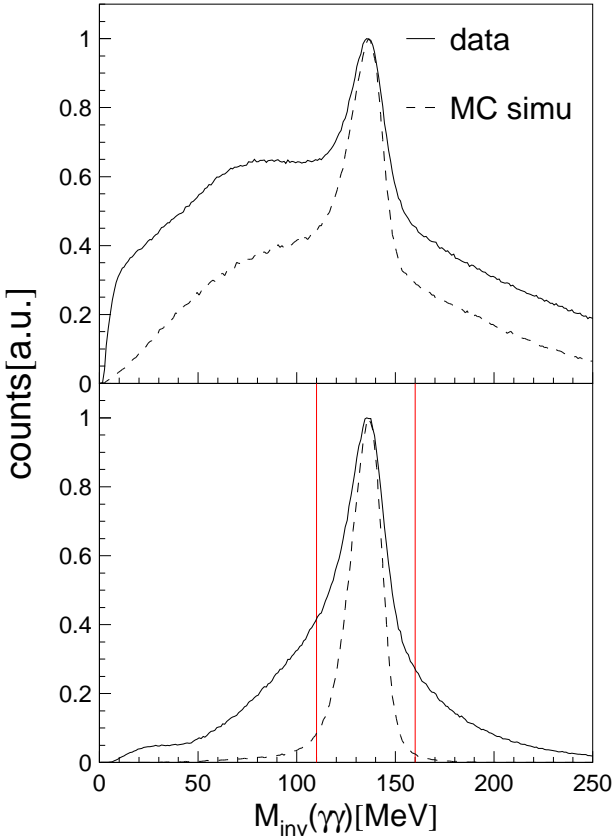}
}}
\caption{Upper part: Invariant mass of photon pairs from events with  
six photons. All possible disjunct combinations of photons are included.
Solid line: data, dashed line: Monte Carlo simulation of 
$\eta\to 3\pi^0\to 6\gamma$. Bottom part: same data samples but only `best' 
combination selected by $\chi^2$ test is plotted (see text). The vertical 
lines indicate the cut applied to select candidates for the 
$\eta\to 3\pi^0$ decay}.
\label{fig:invmas_pi}       
\end{figure}

Even in the presence of intense charged particle background, the analysis 
leads to a very clean photon sample.
The following invariant mass analysis is based on the excellent reproduction
of the shapes of the invariant mass peaks by Monte Carlo simulations.
\begin{figure}[th]
\resizebox{0.50\textwidth}{!}{%
  \includegraphics{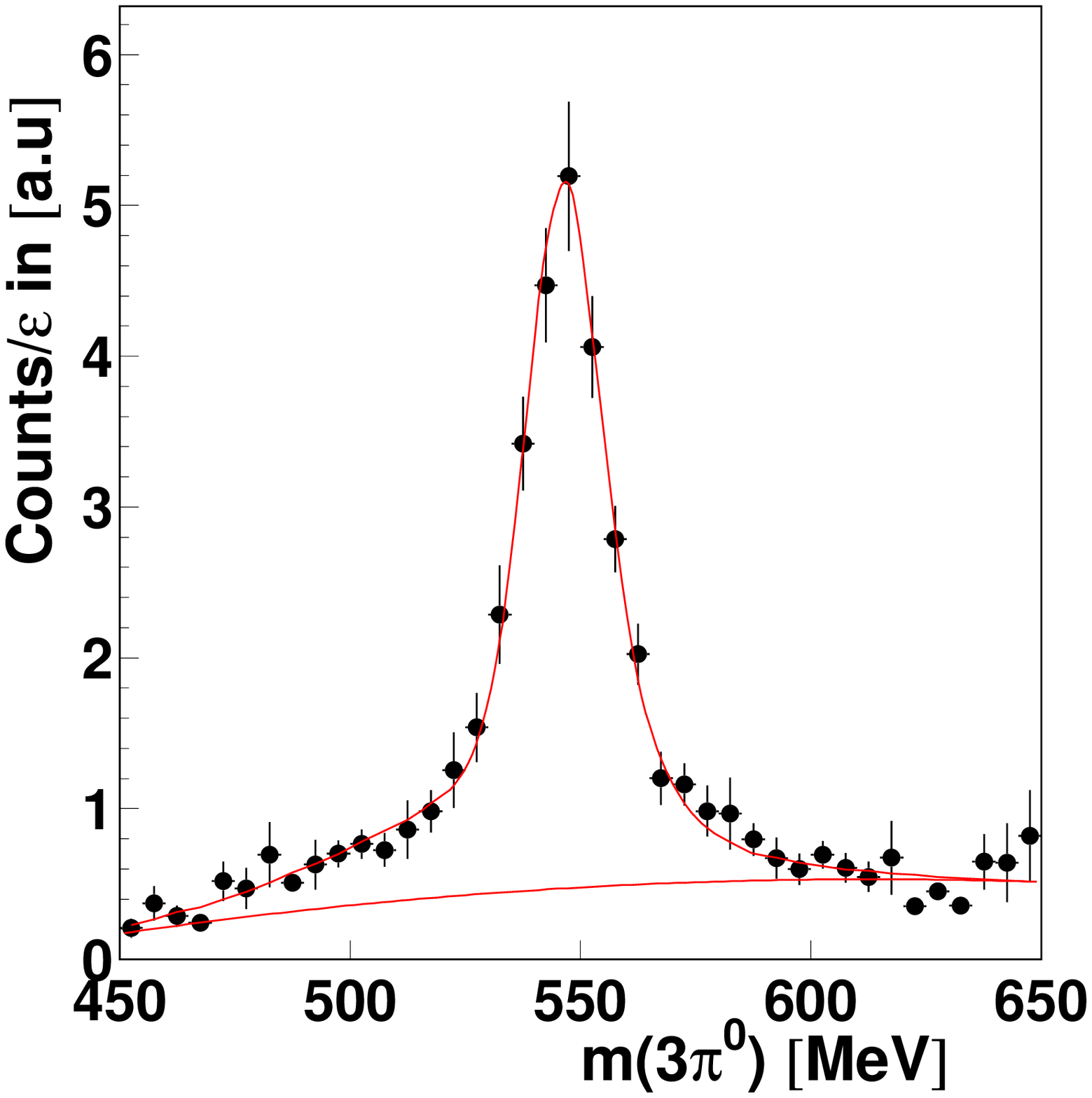}
  \includegraphics{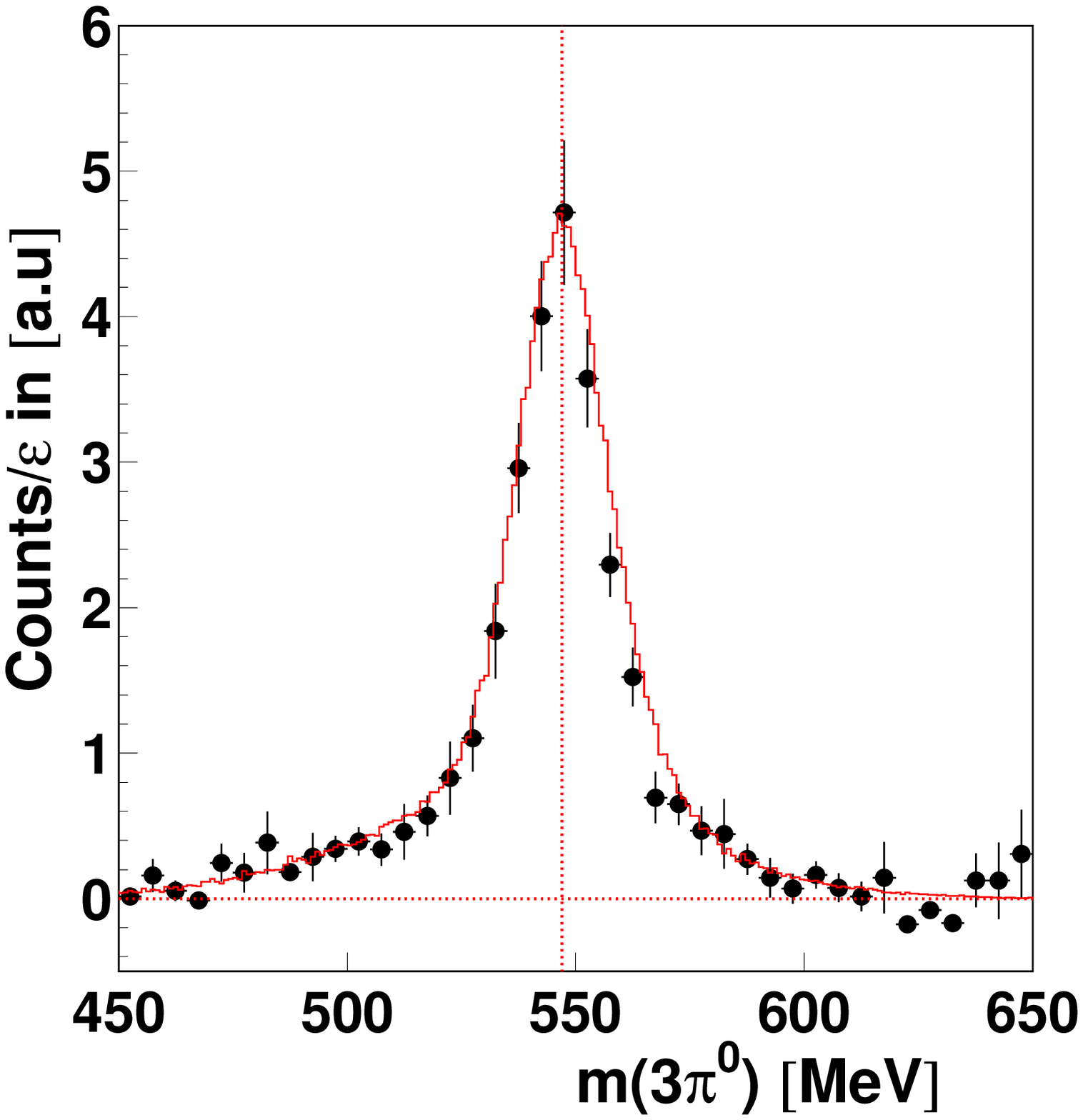}
} 
\resizebox{0.50\textwidth}{!}{%
  \includegraphics{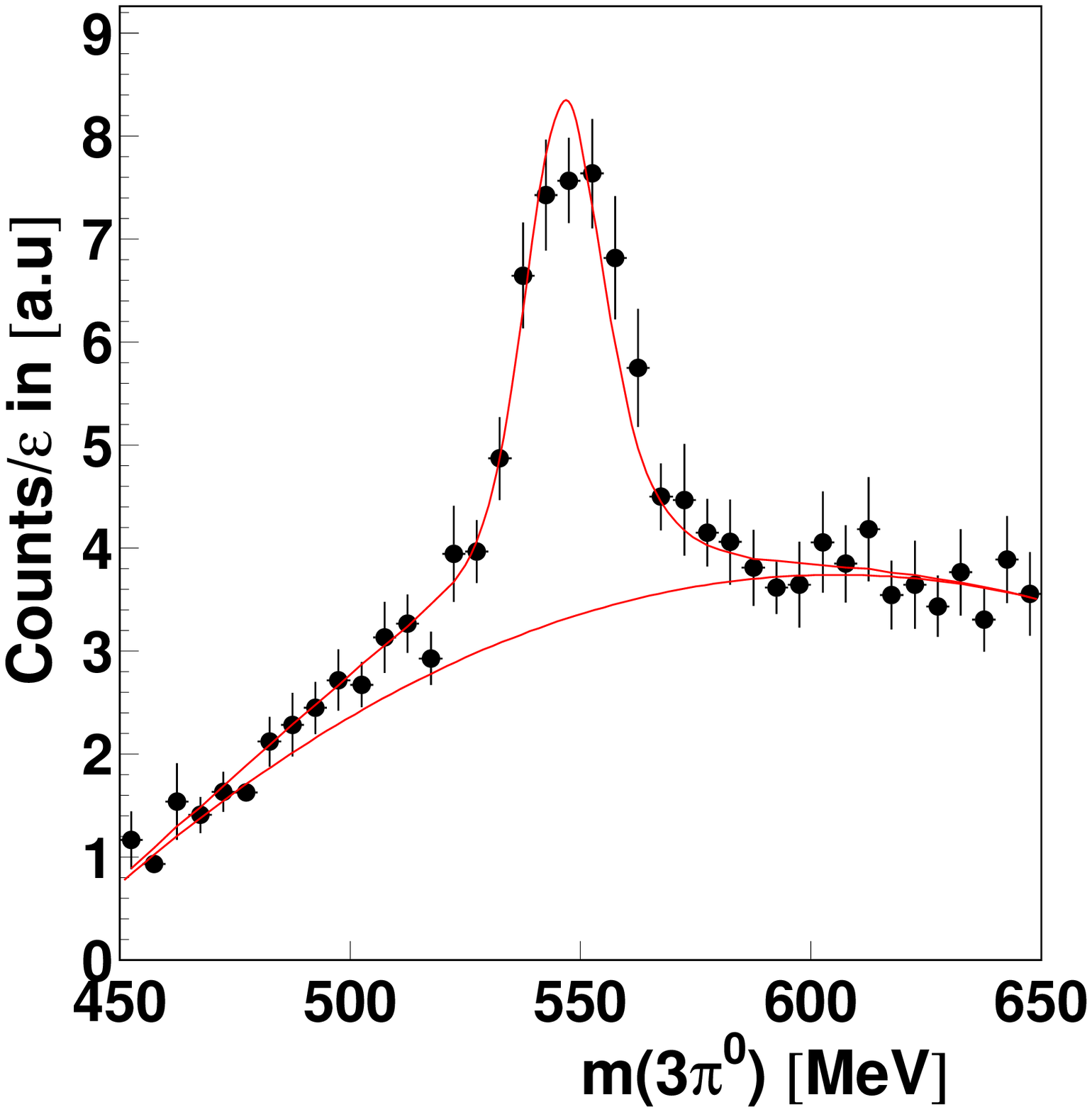}
  \includegraphics{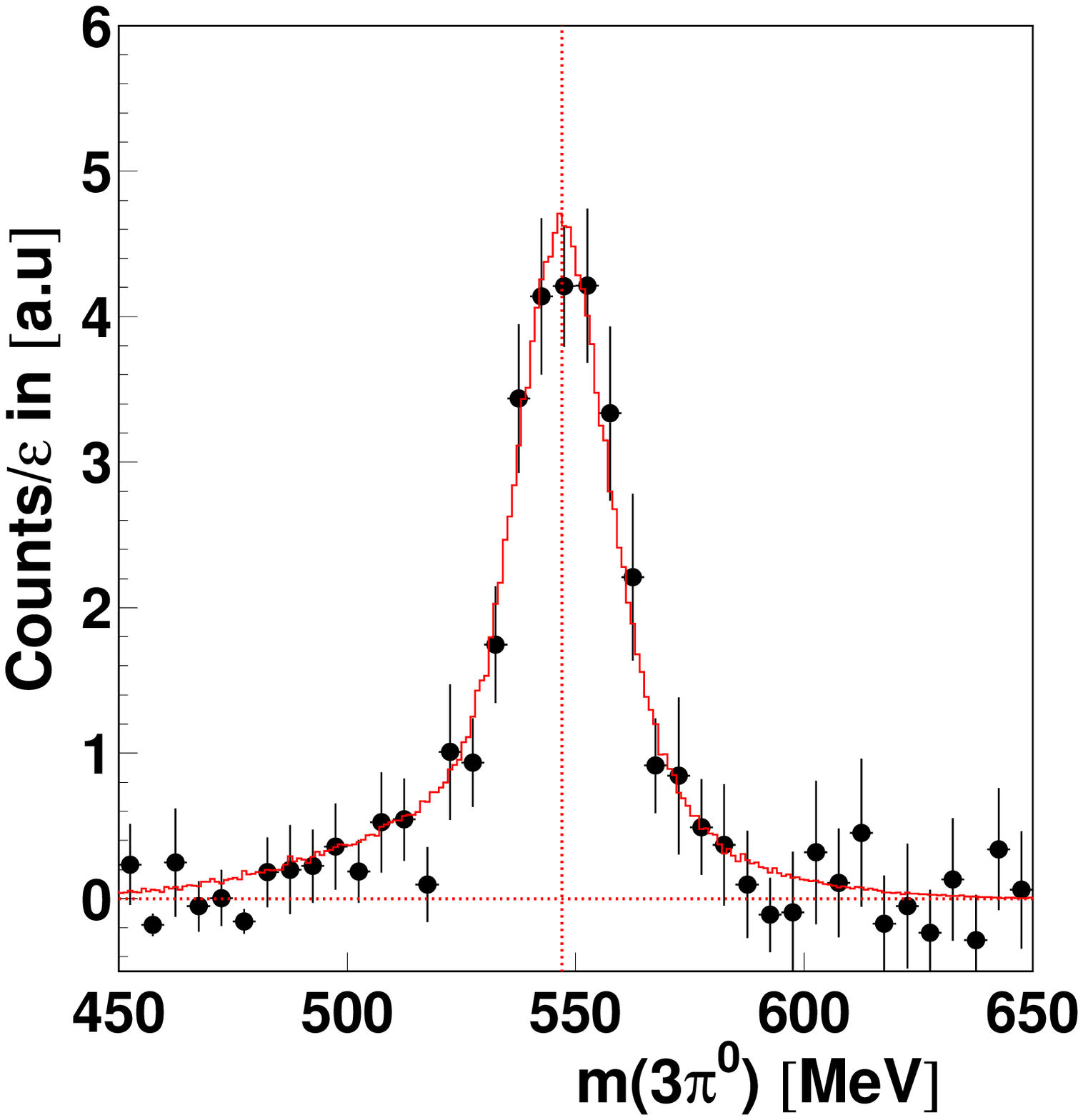}
}  
\resizebox{0.50\textwidth}{!}{%
  \includegraphics{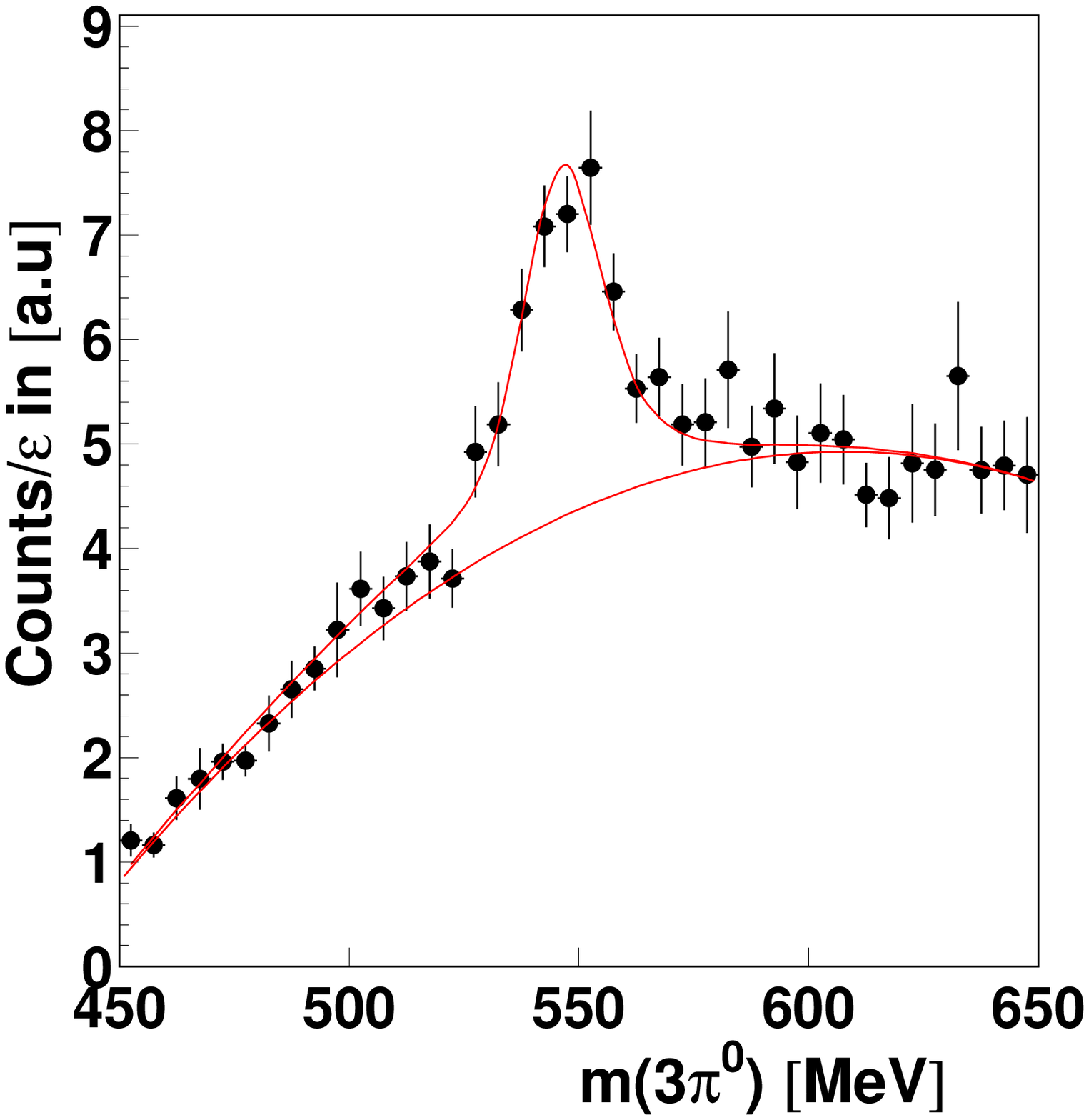}
  \includegraphics{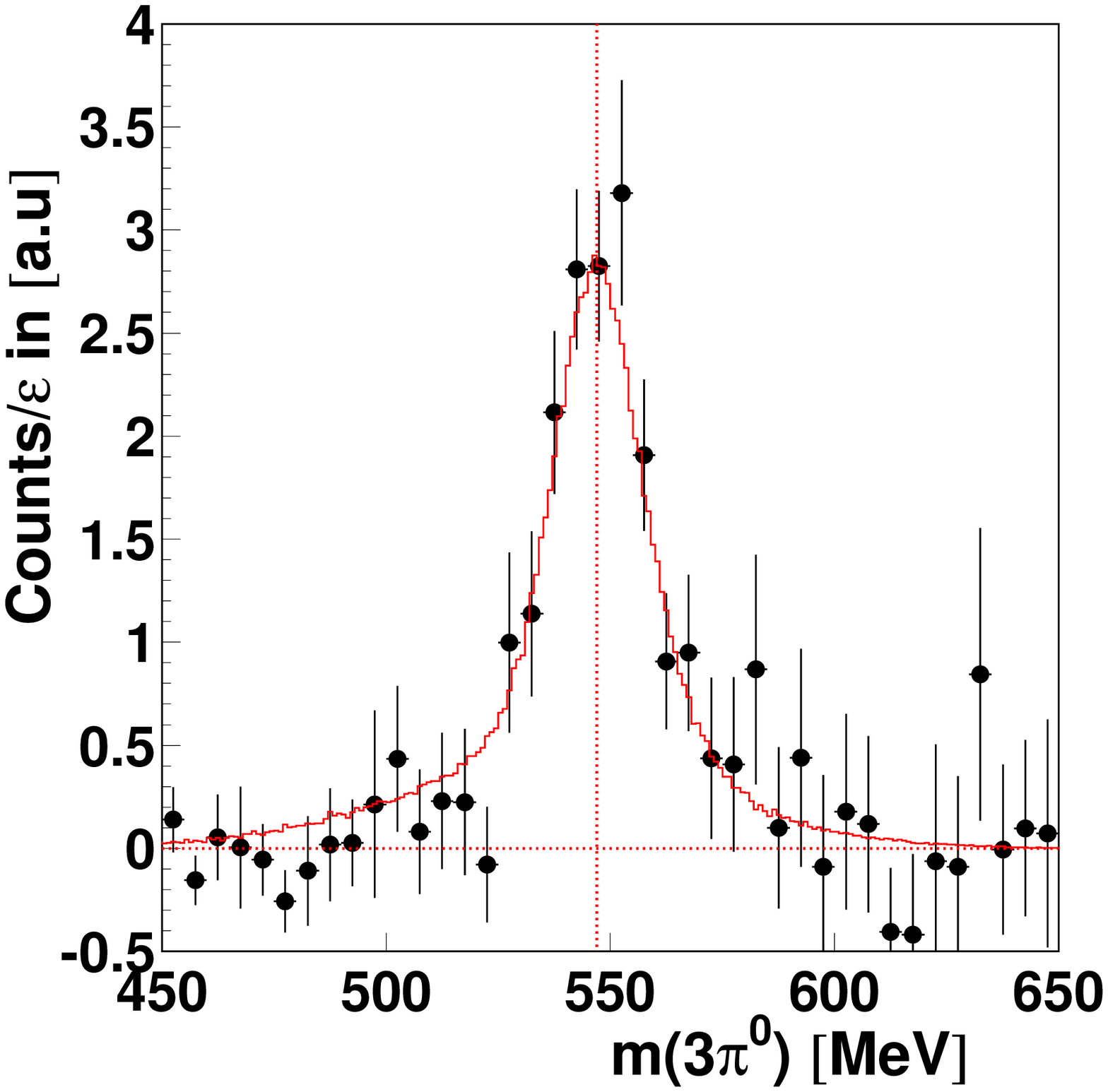}
}
\caption{Identification of $\eta$-mesons via invariant mass. Upper row:
incident photon energies in the range 0.8 - 0.85 GeV, middle: 1.4 - 1.6 GeV,
bottom: 1.85 - 2.15 GeV. Left column: spectra with fitted background and
$\eta$-peak. Right column: background subtracted peaks compared to
Monte Carlo line-shapes. Vertical lines: nominal position of $\eta$ mass peak.
All events have been corrected for detection efficiency
(see text).
}
\label{fig:invmas}       
\end{figure}
In the first step for events with six (or more) 
photons the invariant mass of all possible disjunct photon pairs was 
calculated. In the case of six photons 15 different combinations into 
photon pairs are possible. The resulting two-photon invariant mass spectrum 
is shown in Fig. \ref{fig:invmas} (top part) together with a Monte Carlo
simulation of the $\eta\to 3\pi^0\to 6\gamma$ decay. Among the 15 combinations 
the `best' combination was chosen via a $\chi^2$ analysis which minimizes the 
following expression
\begin{equation}
\chi^2 = \sum_{k=1}^{3}\frac{(m_{\gamma\gamma}(i_k,j_k)-m_{\pi}^0)^2}{(\Delta
m_{\gamma\gamma}(i_k,j_k))^2}
\end{equation}
where $i_1,...,i_3,j_1,...j_3$ represents a permutation of 1,...,6,
$m_{\pi}^0$ is the  pion mass, and the $m_{\gamma\gamma}$ 
are the invariant
masses of the photon pairs with their uncertainties $\Delta m_{\gamma\gamma}$.
The resulting spectra for data and simulation are shown in Fig.
\ref{fig:invmas} (bottom part). In case of the simulation, where the background
in the upper part of the figure is only of combinatorial nature, a clean,
background-free $\pi^0$ invariant mass peak is recovered by this procedure.
For the data, some background from other reactions remains. Due to the selection
procedure of the `best' combination this background is also concentrated around
the peak region. In the next step, events were selected where all
three two-photon invariant masses of the best combination are lying between
110 MeV and 160 MeV (indicated by the vertical lines in fig. \ref{fig:invmas}). 
This cut is motivated by the shape of the simulated invariant mass cut and
removes only a small fraction of `true' events, which is determined from
the simulation and taken into account for the extraction of the cross section.

The nominal mass of the pion was then used as a constraint to improve the
experimental resolution. Since the angular resolution of the detector for 
photons is much better than the energy resolution, it was not necessary to use 
a kinematic fitting procedure. Instead, only the photon energies were 
re-calculated from
\begin{equation}
E'_{1,2}=E_{1,2}\frac{m_{\pi^0}}{m_{\gamma\gamma}}
\end{equation}   
where $E_{1,2}$ are the  measured photon energies, $E'_{1,2}$ the re-calculated,
$m_{\pi^0}$ is the nominal $\pi^0$ mass, and $m_{\gamma\gamma}$ the measured
invariant mass. This procedure was applied to all three photon pairs combining
to pions. In the last step, the invariant mass spectrum of the
six photons is build from their four-vectors, using the measured angles and the 
re-calculated energies. Typical results for the $\eta$
invariant mass peak for different incident photon energies are show in 
Fig. \ref{fig:invmas}.
The spectra can be fitted by a simple
polynomial background (polynomial of second degree) and the line shape of the 
invariant mass peaks generated
from a Monte Carlo simulation with the GEANT3 package \cite{Brun_86}.
For the fit only the three parameters of the background polynomial
and the amplitude of the simulated response function
were varied. Background subtracted spectra compared to the simulated 
line-shape are also shown in Fig. \ref{fig:invmas}.
The shape of the invariant mass peaks is in excellent agreement with the
results of the Monte Carlo simulation (see Fig. \ref{fig:invmas}), where the 
same analysis procedure was applied. The position of the invariant mass
peak as function of incident photon energy is rather stable, which is partly
due to the fact that the positions of the $\pi^0$ and $\eta$ invariant mass 
peaks have been used in an iterative procedure for the energy calibration
of the detector, and partly due to the re-calculation of the photon energies
from the $\pi^0$ invariant masses. 
For differential cross sections, this procedure must be applied to each bin.
The detection efficiency, as discussed in the next section, was
corrected on an event-by-event basis. Therefore, the fitting was not actually
done on the raw invariant mass distributions but on the corresponding spectra 
with efficiency corrected events (shown in fig. \ref{fig:invmas}).

\begin{figure}[th]
\resizebox{0.50\textwidth}{!}{%
  \includegraphics{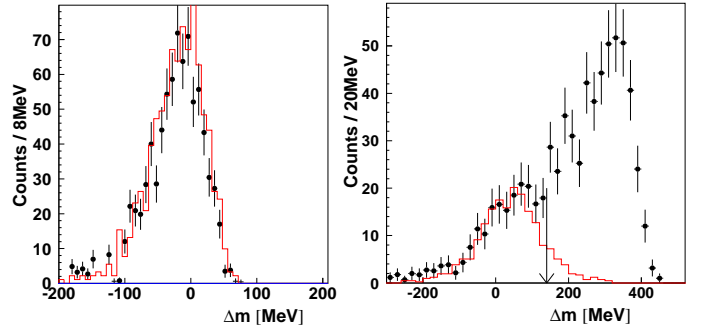}
}
\caption{Missing mass spectra for lead calculated under the assumption of
quasi-free single $\eta$ production. Left hand side: incident photon energies
in the range 0.65 - 0.8 GeV, right hand side: 1.3 - 1.5 GeV. Histograms:
simulated detector response for quasi-free single $\eta$ production. Arrow
indicates cut for this reaction.
}
\label{fig:mismas}       
\end{figure}

With the analysis discussed so far, the fully inclusive reaction
$\gamma A\rightarrow\eta X$ is identified, where $X$ may also
include any (kinematically possible) number of pions. Selecting exclusive,
single $\eta$-production without any further mesons in the final state is
not possible by simply vetoing events with additional clusters in the detector.
Additional charged mesons may go undetected (e.g. along the
beam line), making the suppression incomplete.
In case of the Barrel, charged pions cannot be distinguished
from protons and the condition would falsely suppress events with detected
recoil protons. Therefore, exclusive $\eta$-production can only be identified
via the reaction kinematics.
Here, it is assumed that the reaction occurs quasi-free off a bound nucleon.
The initial momentum of the nucleon is neglected and the missing mass $\Delta m$
of the reaction is calculated from the nucleon mass $m_N$, the energy of the
incident photon $E_b$, and the energies and momenta $E_{\gamma_i}$ and
$\vec{P_{\gamma_i}}$ of the six decay photons:
\begin{equation}
\Delta m = \sqrt{(E_b+m_N-\sum_{i=1}^{6}{E_{\gamma_i}})^2
-(\vec{P_b}-\sum_{i=1}^{6}{\vec{P_{\gamma_i}}})^2}- m_N.
\label{eq:mismas}
\end{equation}
The resulting distributions are broadened by Fermi motion, so that a
perfect separation of the different reaction channels is not possible.
Examples of missing mass spectra are shown in fig. \ref{fig:mismas}.
The structures around zero missing mass are related to single $\eta$-production,
the contribution at large positive missing mass, which is only visible at
higher incident photon energies, originates from $\eta\pi$ final states
and secondary production processes like $\gamma N\rightarrow \pi N$,
$\pi N\rightarrow \eta N$. The experimental results are compared to a
simulation of the detector response based on the shape of the missing mass
distributions for single quasi-free $\eta$-production predicted by a BUU
model (see sec. \ref{sec:buu}). The indicated cut was used for the analysis
of single $\eta$-production. This cut does not allow a perfect
separation of single $\eta$-production from the other contributions since the
tails of the different distributions are overlapping. However, since the same
cut was used for the model results (see sec. \ref{chap:buucom}), the comparison
between data and model is not affected.

\section{Determination of cross sections}
The absolute normalization of the measured yields was obtained from the
target densities, the incident photon flux, the
$\eta\rightarrow 3\pi^0\rightarrow 6\gamma$ decay branching ratio
($b_{\eta\rightarrow 6\gamma}$=31.35 \%), and the detection efficiency of the
calorimeter. The measurement of the photon flux is based on the counting
of the deflected electrons in the focal plane detectors of the tagging
spectrometer (see. fig. \ref{fig:tagger}).
The fraction of correlated photons, passing the collimator and
impinging on the target, was determined roughly once per day in a mode
where the trigger is derived from the tagger and the photon flux
(at reduced beam intensity) is measured by a photon counter placed downstream
of the calorimeter. The detector dead-time of approximately 60 \% was
determined with scalers gated by life-time of the experiment and 
spill-time of the accelerator, respectively.

\begin{figure}[hbt]
\resizebox{0.50\textwidth}{!}{%
  \includegraphics{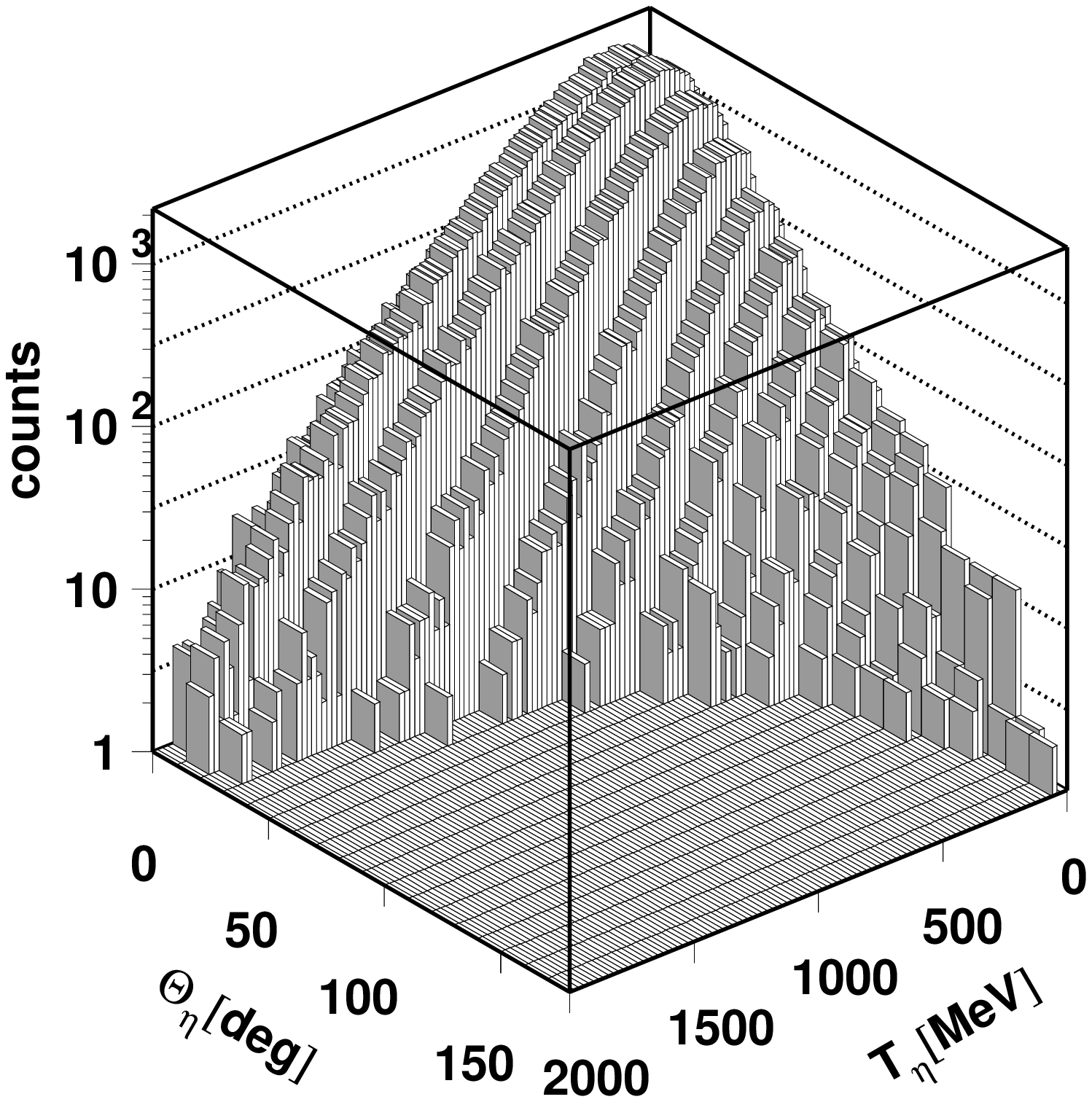}
  \includegraphics{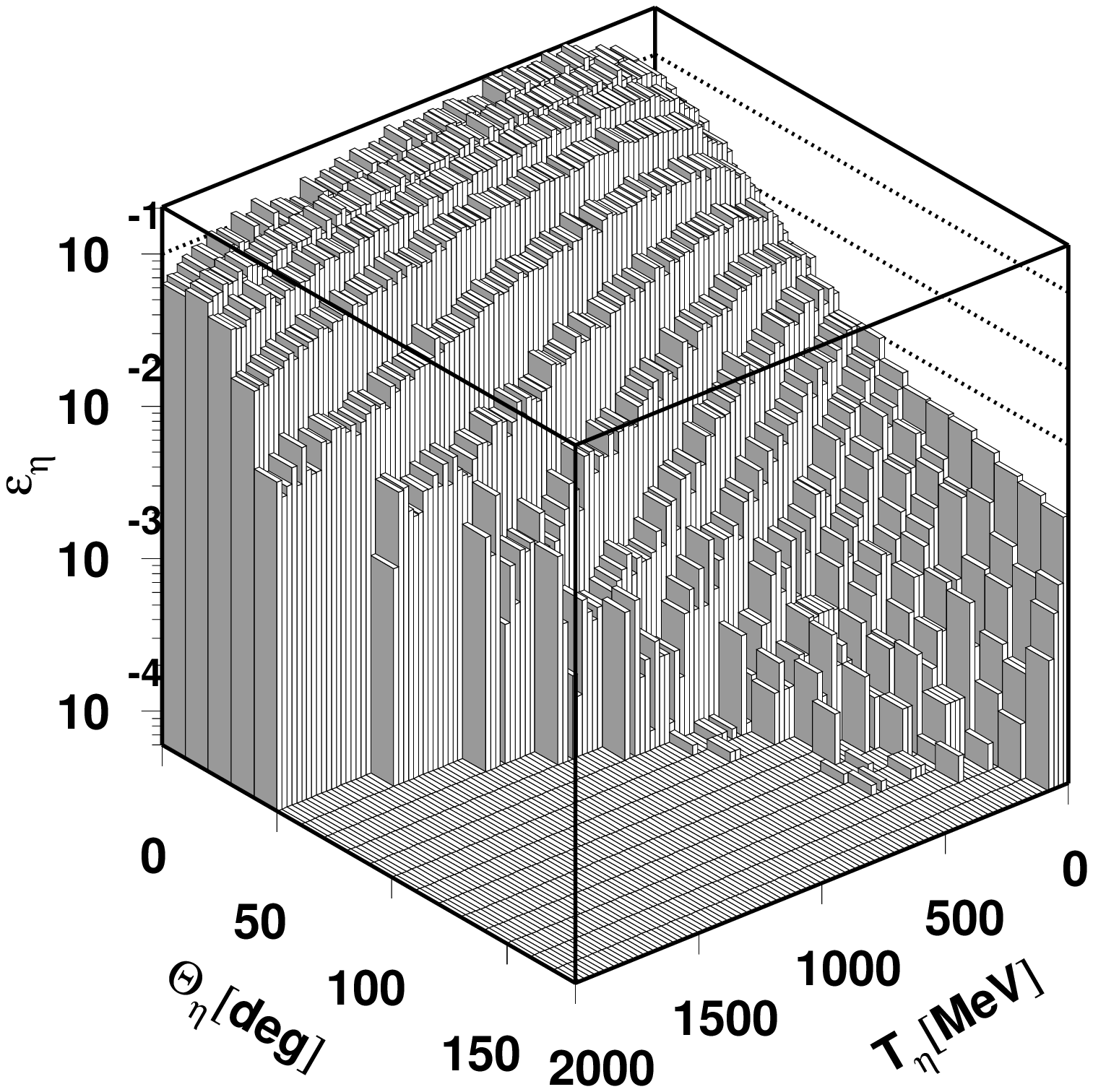}
}
\caption{Left hand side: laboratory angle and kinetic energy distribution
of measured $\eta$-mesons. Right hand side: detection efficiency as function
of the same parameters (regions clearly outside kinematically possible
combinations have not been simulated).}
\label{fig:effi}       
\end{figure}
 
The detection efficiency of the Crystal Barrel/TAPS calorimeter has been
determined by Monte Carlo simulations using the GEANT3 package \cite{Brun_86}.
The simulation includes all relevant properties of the experimental setup like
geometrical acceptance, trigger efficiency, detection efficiency, and  
analysis cuts. The branching ratio for the $\eta\rightarrow 6\gamma$ decay
is not included in the efficiency. 
The $\eta$ mesons are produced in many different final states and
reaction types: single $\eta$, $\eta\pi$, re-scattered $\eta$ mesons, and 
$\eta$ mesons from secondary reactions in the nucleus. All processes are 
additionally complicated by the momentum distribution of the bound nucleons. 
Consequently, the correlation between momentum and emission angle of the mesons 
is a priori not known. Therefore, a reliable model is not available for an 
event generator for the Monte Carlo simulations.
Instead, the detection efficiency has been determined from the simulations
as a function of the laboratory polar angle and laboratory
kinetic energy of the $\eta$-mesons, which are measured quantities.
It was then applied on an event-by-event basis to the data.
The method is described in more detail in \cite{Krusche_04} for inclusive
$\pi^0$ production off nuclei.
In this way, a model independent detection efficiency correction is achieved
as long as the efficiency does not vanish for any kinematically possible
combination of $\eta$ polar angle and kinetic energy.
This is demonstrated in fig. \ref{fig:effi} where the correlation between
angles and energies of the measured $\eta$ mesons is compared to the simulated
detection efficiency. The absolute values of the detection efficiency are not
large since the first level trigger was only sensitive to photons in the TAPS 
forward wall (see sec. \ref{sec:setup}). Even for the six-photon decay 
of the $\eta$ the efficiency of this trigger condition is small, in particular 
for $\eta$ mesons emitted at large polar angles. However, it is obvious from 
the figure, that the entire phase space of kinematically possible combinations 
is covered by non-vanishing detection efficiency, so that for the 
determination of cross sections no extrapolations had to be done.

\section{Systematic uncertainties and comparison to previous results}

The main sources for systematic uncertainties are related to the background
level in the $\eta$-invariant mass spectra, the simulation of the detection
efficiency, and the determination of the incident photon flux.
Other uncertainties like e.g. the surface thickness of the solid targets
(better than 1 \%) are comparably negligible.

\begin{figure}[th]
\resizebox{0.50\textwidth}{!}{%
  \includegraphics{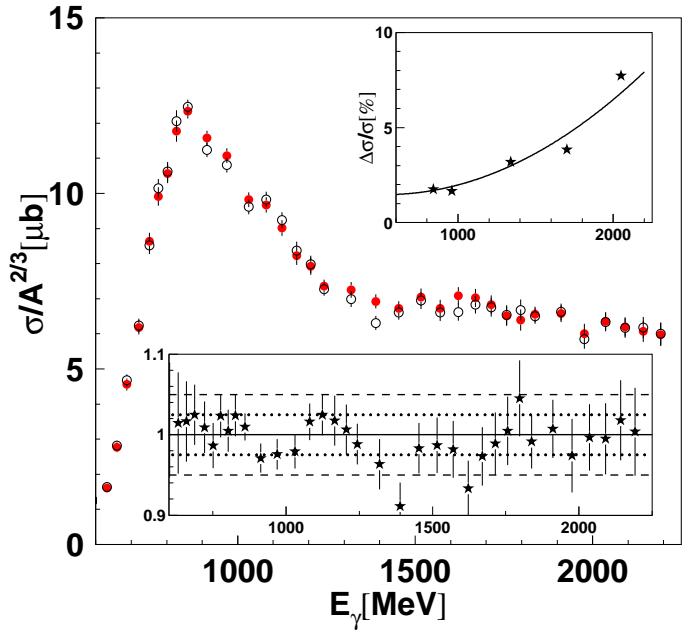}
 }

\caption{Typical systematic effects (carbon data). Main plot: total cross
section from detection efficiency with 10$^o$ binning in $\theta_{\eta}$
(filled (red) circles) compared to 2$^o$ binning (open (black) circles).
Bottom insert: ratio of cross sections. Top insert: typical systematic
uncertainty related to background shape in $\eta$-invariant mass spectra 
(see text).}
\label{fig:systy}       
\end{figure}

\begin{figure}[th]
\resizebox{0.50\textwidth}{!}{%
  \includegraphics{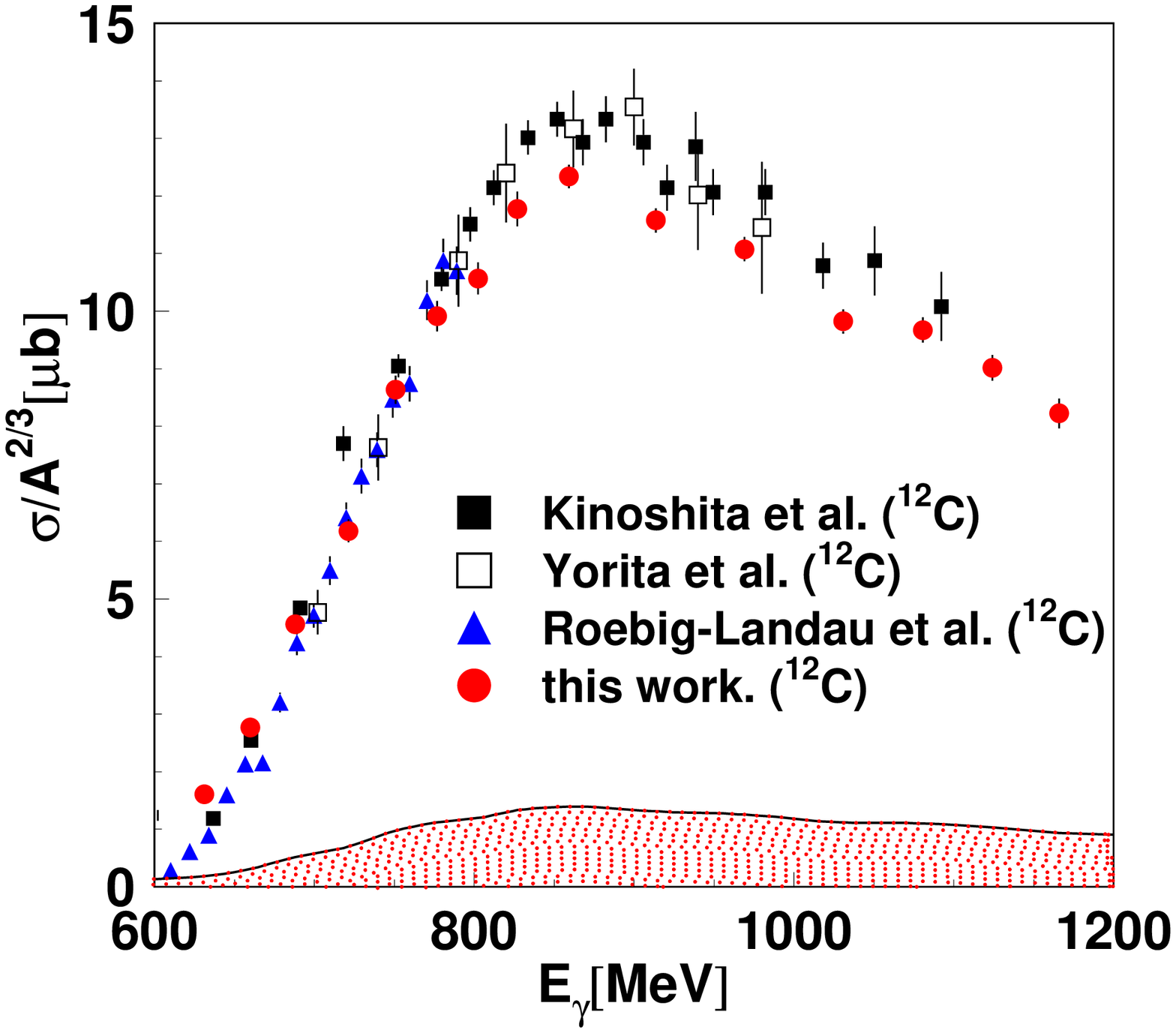}
 }
 \resizebox{0.50\textwidth}{!}{%
  \includegraphics{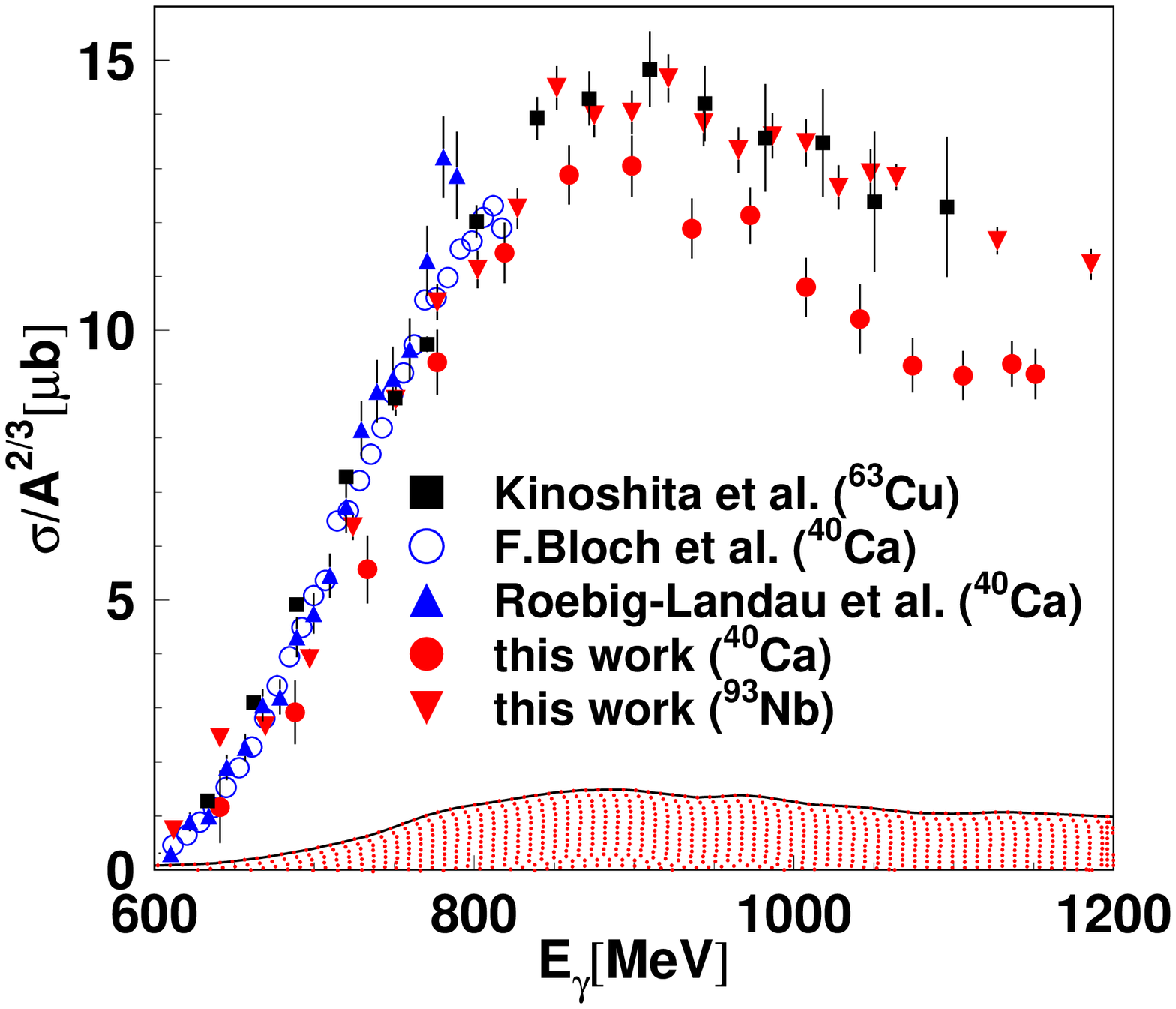}
}
\caption{Comparison with earlier results. Upper part: $^{12}$C from Mainz
\cite{Roebig_96}, Tohoku \cite{Kinoshita_06} and this work. Bottom part:
$^{40}$Ca from Mainz \cite{Bloch_07}, $^{63}$Cu from Tohoku \cite{Kinoshita_06},
and $^{40}$Ca and $^{93}$Nb this work. The error bars of the present data
include only the statistical errors, the shaded bands at the bottom indicate
the total systematic uncertainties for $^{12}$C (top) and $^{40}$Ca
(bottom).}
\label{fig:previous}       
\end{figure}

As discussed in sec. \ref{sec:datana}, the background level beneath the 
$\eta$ signal has been determind by fitting the amplitude of the simulated 
shape of the invariant mass peak and the three parameters of a background
polynomial. The resulting background subtracted signal, shown in fig 
\ref{fig:invmas}, agrees well with the simulated line shape. The 
systematic uncertainty of this procedure was studied by a variation of the 
fitted range in the spectra, giving rise to systematic deviations in the
background shape. Typical systematic variations of the fitted peak amplitude
(see fig. \ref{fig:systy}, upper insert) range from 2\% at low incident photon 
energies to 8\% at the highest incident photon energies.

Since the detection efficiency could be simulated without any assumptions 
about the angular and energy distributions of the mesons, the corresponding
uncertainties are small, estimated at the 5 \% level. They are mainly due to
the exact representation of thresholds, shower development, absorbing inactive 
material in the detector, and the exact target and beam positions in the 
GEANT simulations. 
The stability of the detection efficiency correction has been checked by a
variation of the bin size of the simulated efficiency by a factor of five 
(Fig. \ref{fig:effi} shows the coarsest binning in angle). 
Figure \ref{fig:systy} shows a comparison of the total cross section for
carbon constructed with bin sizes of 10$^o$ and 2$^o$. Most data points from 
the two analyses agree within $\pm$2.5\% and the fluctuations are mostly
within statistical uncertainties (see fig. \ref{fig:systy}, bottom insert).

The photon flux involves the determination of the rate of scattered electrons
and the measurement of the tagging efficiency.
The uncertainty could be estimated from the measurement with the deuteron target.
In this case, different incident photon spectra have been used.
About half of the data were measured with an electron beam energy of
2.6~GeV and a linearly polarized photon beam (coherent Bremsstrahlung from
a diamond lattice, see \cite{Elsner_07} for details) with the polarization
maximum around 1~GeV. The other data were measured with an electron beam energy
of 3.2~GeV and unpolarized photon beam. Consequently, the shapes of the
incident photon spectra were different, due to the coherent peak.
Due to the different beam energies, scattered electrons corresponding to the
same photon energy have been registered in different sections of the tagging
spectrometer. After the flux correction, the cross sections for $\eta$
production extracted from the two runs agreed within $\pm$10 \%, taken as the
typical systematic uncertainty. 

The flux uncertainty is identical for all nuclei (same settings of incident
photon energy, beam intensity, and parameters of the tagging system
for all nuclei) and thus does not influence $A$-dependent properties.
It must only be accounted for in the comparison to model results or 
data from other experiments. For this case, all three systematic effects 
have been added in quadrature. The systematic uncertainties of the detection 
efficiency and the background subtraction are also not completely independent
for the four nuclei. However, for an estimate of systematic uncertainties of
scaling properties we have made the most pessimistic assumption that this
effects vary independently. 

The total cross section data for inclusive $\eta$ production are compared in
fig. \ref{fig:previous} to previous results below 1.2~GeV. Carbon and Calcium
data are available from Mainz below 0.8~GeV \cite{Roebig_96,Bloch_07},
Carbon data from KEK \cite{Yorita_00} and Carbon and Copper data from
Tohoku \cite{Kinoshita_06} below 1.2~GeV. In the Carbon case, 
the KEK and Tohoku data are systematically higher by roughly 10 \%. For the 
heavier nuclei, a direct comparison between the Tohoku results and the present 
data is not possible, since different target nuclei have been investigated. 
At incident photon energies below 800 MeV, all data scale like $A^{2/3}$.
At higher incident photon energies, the $A^{2/3}$ scaling does not hold 
anymore, the present Calcium data are clearly lower than the present Niobium 
data (see also fig. \ref{fig:total_incl}).
From this behavior, one expects that the scaled cross section for Copper
lies in between the Calcium and Niobium results. However, the Tohoku Copper
results fall on top of our Niobium data, indicating that also for the heavier
nuclei the Tohoku results are systematically higher by roughly 10 \%.
However, the energy dependence of the excitation functions for the present
data and the KEK and Tohoku results is very similar. Rather good agreement is
found when all data are re-scaled by 10 \%, which is within their systematic
uncertainty. 

\section{The BUU-model calculations}
\label{sec:buu}

The detailed interpretation of the experimental results is only possible
via a comparison to model calculations which incorporate effects like
nuclear Fermi motion, Pauli-blocking of final states, and in particular
the propagation and absorption of mesons and nucleon resonances in nuclear
matter. Results obtained in the framework of the BUU transport model
for photon induced reactions as discussed in detail in
\cite{Hombach_95,Effenberger_97,Lehr_00} have been used.
The model is based on the BUU equation:
\begin{eqnarray}
  \left({\partial\over\partial t}+\vec{\nabla}_p H\cdot\vec{\nabla}_r
  -\vec{\nabla}_r H\cdot\vec{\nabla}_p\right)F_i(\vec r,\vec p,\mu;t)=\nonumber
\\
= I_{coll}[F_N,F_\pi,F_\eta,F_{N^{\star}},F_{\Delta},...]
\end{eqnarray}
which describes the space-time evolution
($\vec{r}$: space coordinate, $\vec{p}$: momentum) of the spectral phases
space density $F_{i}$ of an ensemble of interacting particles
$i$=$N, N^{\star}, \Delta, \pi, \eta,...$
in nuclear matter from the moment of their creation to their absorption or
escape through the outer boundaries of the nucleus. The left hand side of the
equation - the Vlasov term - describes the propagation of the particles under
the influence of a Hamilton function $H$. It contains information about
energy, mass, self-energy (mean field) of the particle and a term that drives
back an off-shell particle to its mass shell. Explicitly, it can be written as
\begin{equation}
H = \sqrt{(\mu + S)^2+p^2}
\end{equation}
with the particle mass $\mu$ and a scalar potential $S$ for baryons 
\cite{Lehr_00}.
The right hand side of the BUU equation - the collision integral -
describes particle production and absorption. It consists of a gain and a loss
term for the phase space density $F_{i}$, accounting for interactions between
the particles beyond the mean-field potential.

The constituents of the nucleus are defined as `test nucleons'
and follow a Woods-Saxon density distribution

\begin{equation}
\label{eq462}
\rho (r)
=
\frac{\rho_{o}}{1+e^{(r-R)/a}}\;\;\;,
\end{equation}

where the nuclear radius is related to the nucleus mass
via $R=1.124A^{1/3}$ fm and $a=(0.0244\;A^{1/3}+0.2864)fm$.
The momentum distribution is described within the Fermi gas approach.

\begin{equation}
\label{eq463}
p_F(r)
=
\left(\frac{3\pi^2}{2}\rho(r)\right)^{1/3}.
\end{equation}

The elementary $\eta$ cross sections off protons and neutrons
are included in this model. The produced resonances and mesons
propagate in the nucleus and can be scattered, absorbed or decay.
The different reaction probabilities are either
fitted to experimental data or calculated. They are incorporated
into the model by the collision term and may interact according
to the geometrical condition that the distance between the two particles
is smaller than the impact parameter $b_{c} = \sqrt{\sigma / \pi}$ where
$\sigma$ is the reaction cross section.

\section{Results}

\begin{figure}[htb]
\centerline{\resizebox{0.44\textwidth}{!}{%
  \includegraphics{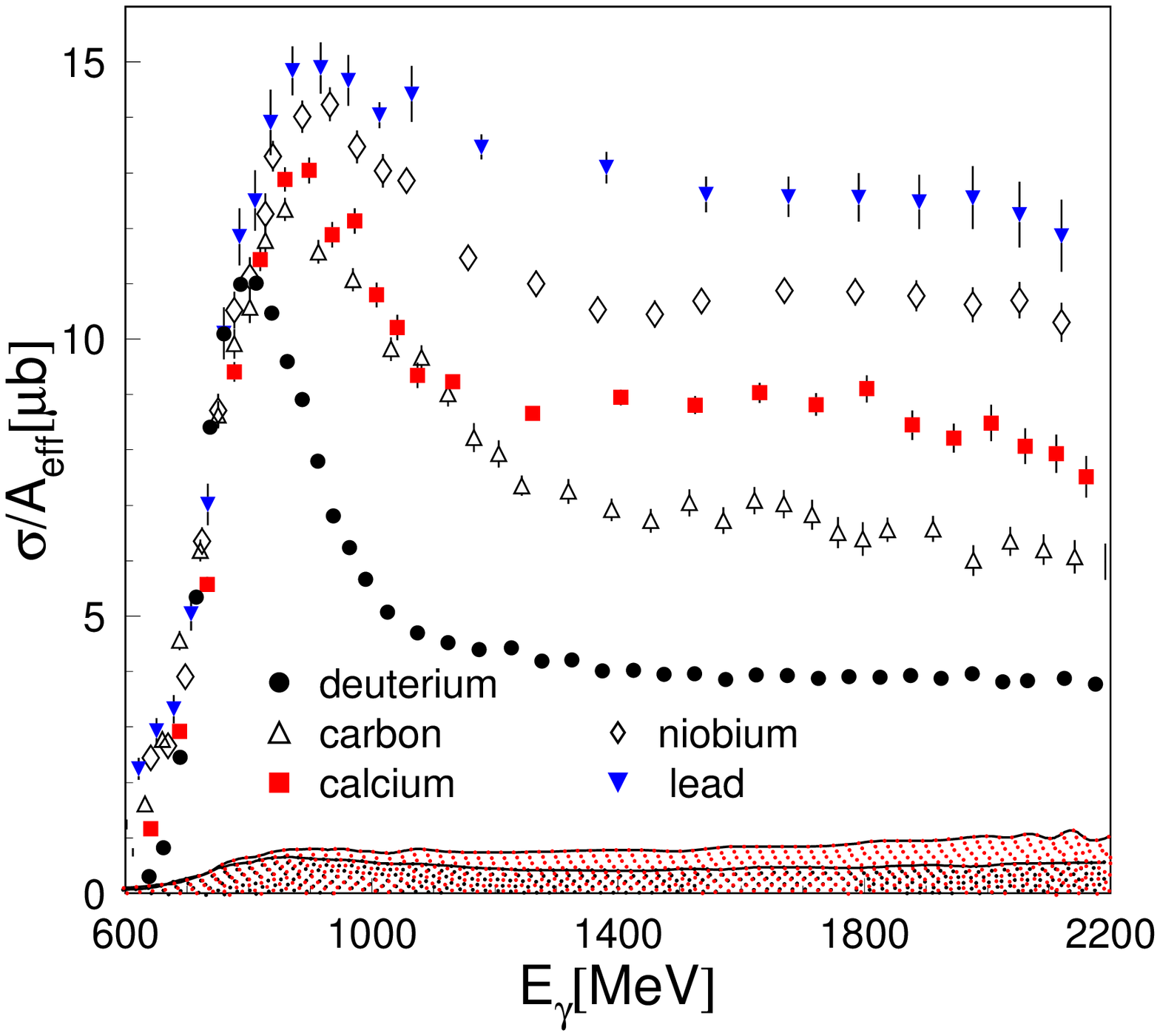}
}}
\centerline{
\resizebox{0.44\textwidth}{!}{%
  \includegraphics{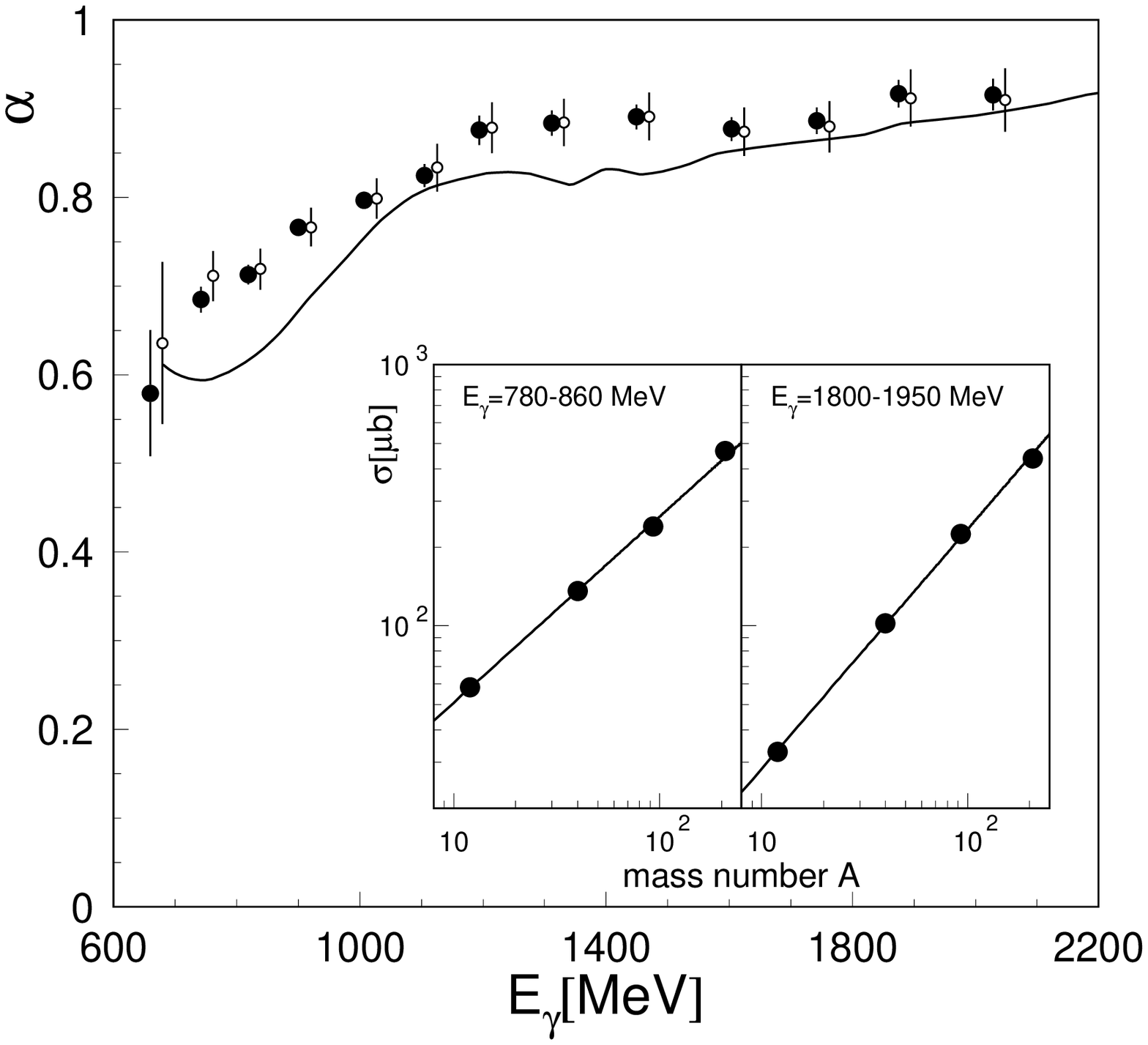}
}}
\caption{Upper part: Total fully inclusive $\eta$ cross section normalized by
$A_{eff}=A^{2/3}$ for $A=12,40,93,208$ and $A_{eff}=2$ for the deuteron.
Error bars are statistical, the shaded bands at the bottom show the
systematic uncertainties for carbon (lower double shaded band) and lead
(upper band) excluding the 10\% flux normalization uncertainty.
Bottom part: Scaling coefficient $\alpha$ (see text) as function of incident 
photon energy. Black circles: results if only statistical uncertainties are
considered. Slightly displaced open symbols: including systematic uncertainties
except 10\% overall normalization. Curve: result from BUU model.
The two inserts show the scaling with mass number for two typical
ranges of incident photon energy.
}
\label{fig:total_incl}       
\end{figure}

The total inclusive $\eta$ production cross sections (i.a. for the reaction
$\gamma A\rightarrow \eta X$, without any condition for $X$) are summarized
in fig. \ref{fig:total_incl}. At incident photon energies below
$\approx$800 MeV the cross sections scale with $A^{2/3}$ ($A$ atomic 
mass number) for the heavier nuclei and agree with the average nucleon cross
section (deuteron cross section scaled by a factor of 2). In the following
$A_{eff}$ means $A$=2 for the deuteron and $A^{2/3}$ for all other nuclei.
This behavior, which
indicates strong absorption of the mesons, was already found in
\cite{Krusche_04} for $\eta$, $\pi^0$, and double $\pi$ photoproduction in
the same energy range. However, at higher energies, the cross sections behave
completely differently and scale almost with the mass number. 

This is shown more quantitatively in fig. \ref{fig:total_incl} (bottom) where 
the scaling coefficient $\alpha$ obtained from a fit of
\begin{equation}
\sigma(A)\propto A^{\alpha}
\end{equation}
is plotted versus the incident photon energy. It would be tempting to argue that
the rise of the scaling coefficient simply reflects a decrease of the absorption
cross section with increasing kinetic $\eta$-energy, since the most efficient 
absorption process is s-wave excitation of the S$_{11}$(1535) resonance. However,
the situation is not that simple.
For the further discussion, we must keep in mind, that the scaling is not only 
influenced by the absorption cross section of the $\eta$-mesons, but may also 
reflect $A$-dependent differences of their production. In the most simple case 
of quasi-free single-meson production, the production before FSI will scale with 
$A$ and then a scaling of the observed meson rates with $\alpha$=2/3 indicates 
strong FSI, while a scaling with $A$ indicates transparent nuclear matter.
In this case, the scaling coefficient will only depend on the kinetic energy $T$ 
of the mesons (energy dependent absorption cross section).  

\begin{figure}[hb]
\resizebox{0.45\textwidth}{!}{%
\includegraphics{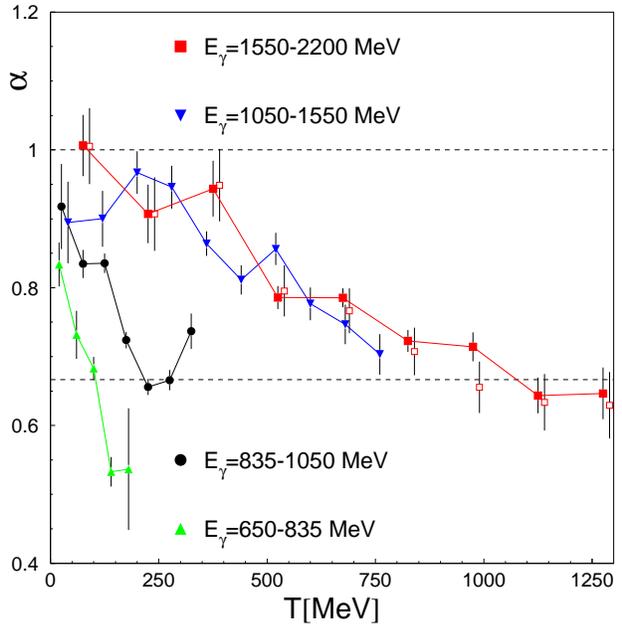}
}
\caption{Evolution of the scaling factor $\alpha$ with the kinetic energy for
different ranges of incident photon energies. The open, slightly displaced
symbols show for the highest incident photon energy the result when 
systematic uncertainties except the 10\% flux normalization are included. 
}
\label{fig:scale}
\end{figure}

\begin{figure}[thb]
\resizebox{0.50\textwidth}{!}{%
\includegraphics{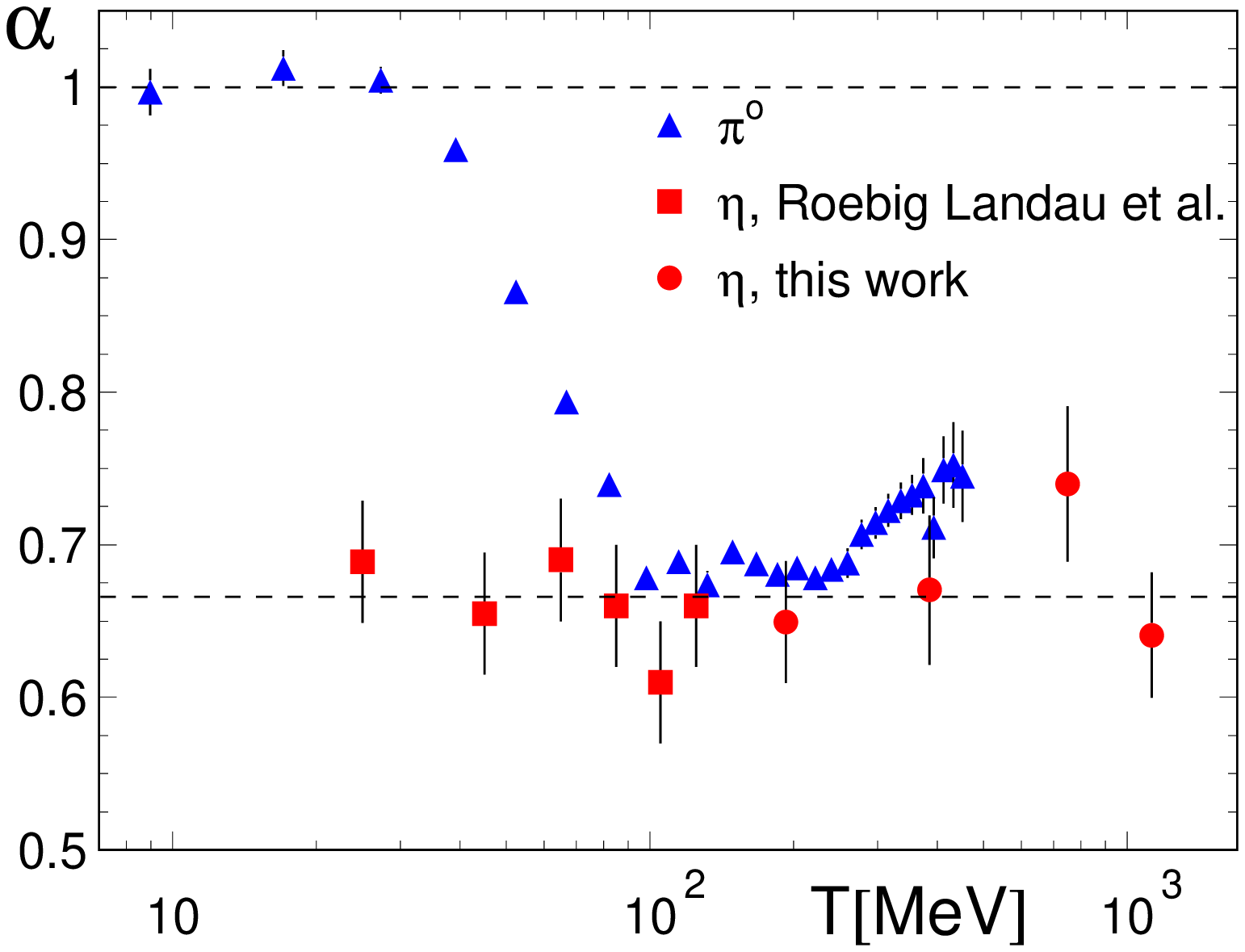}
}
\caption{Evolution of the scaling factor $\alpha$ with the kinetic energy for
$\pi^0$ \cite{Krusche_04} and $\eta$ mesons (low energy $\eta$ data from
\cite{Roebig_96}.
}
\label{fig:eta_pio}
\end{figure}

However, the data show a different behavior. As shown in Fig. \ref{fig:scale}
for fixed values of $T$ the scaling is dependent on the incident photon energy 
$E_{\gamma}$. Furthermore for fixed incident photon energy the coefficients
drop from values close to unity for small $T$ to roughly 2/3 for the largest
$T$ possible at that incident photon energy. This is exactly the opposite
of what one would expect when the behavior of the scaling coefficients would
be dominated by the s-wave absorption into the S$_{11}$.
However, the observed behavior can arise when the production rates of the 
mesons before absorption do not scale with $A$. Problematic in 
this respect are side-feeding contributions from secondary production processes like 
$\gamma N\rightarrow\pi N$, $\pi N\rightarrow\eta N$, and FSI processes
that modify the observed energy distribution of the $\eta$ mesons.
They may completely obscure the effects related to $\eta$-absorption, 
since they may strongly increase with mass number. The BUU-model simulations 
predict, that both contributions are substantial
(see sec. \ref{chap:buucom}). 

However, due to energy and momentum conservation, secondary production processes 
(as well as $\eta\pi$ final states) cannot contribute at the kinematical limit 
(i.e. at maximum $T$ for given $E_{\gamma}$), but will produce $\eta$ mesons 
with smaller kinetic energies (some energy is carried away by the additional 
nucleon(s) involved in the secondary reactions). 
Therefore, it is possible to extract the $\eta N$-absorption cross section 
from the scaling behavior in this regime.
For this purpose, the scaling factors $\alpha$ have been fitted
for the high energy end of the $T$ distributions for different incident
photon energies using the condition
\begin{equation}
T > (E_{\gamma} - m_{\eta})/2
\end{equation}
where $E_{\gamma}$ is the incident photon energy and $m_{\eta}$ the mass of
the $\eta$ mesons. The result as function of $\eta$ kinetic energy is compared
in fig. \ref{fig:eta_pio} to the previous results for low-energy $\eta$ mesons
\cite{Roebig_96} and for $\pi^0$ mesons \cite{Krusche_04}.
The somewhat surprising result is that for $\eta$ mesons the scaling
coefficient $\alpha$ is almost constant at 2/3 over a large range of $\eta$
kinetic energy, indicating strong absorption independent of kinetic
energy. In the case of pions, the absorption is expectedly very weak for 
kinetic energies, which are too low to excite the $\Delta$ resonance.
The pions escape from the nucleus and $\alpha$ (pions) is one. The scaling
factor reaches 2/3 in the $\Delta$ regime and then seems to slowly
increase again.
\begin{figure}[bth]
\resizebox{0.50\textwidth}{!}{%
\includegraphics{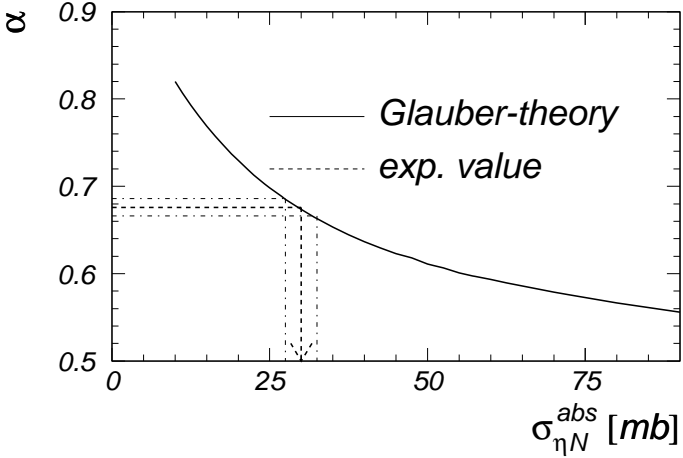}
}
\caption{Dependence of the scaling coefficient $\alpha$ on the $\eta N$
absorption cross section \cite{Roebig_96}.
}
\label{fig:glauber}
\end{figure}
The large absorption cross section for $\eta$ mesons at small kinetic energies
was expected since this is the excitation region of the S$_{11}$ resonance with
a strong coupling to the $N\eta$ channel. Unexpectedly, a decrease of the 
absorption cross section is not observed, even at kinetic energies far above 
this range.
The corresponding $\eta N$ absorption cross section $\sigma^{abs}_{\eta N}$ can 
be deduced from the results of a Glauber model calculation discussed in 
\cite{Roebig_96}. The model is based on the assumption, that secondary production 
processes play no role, which has been assured, as discussed above, by the choice 
of the kinematical conditions. The dependence of the scaling coefficient 
$\alpha$ on $\sigma^{abs}_{\eta N}$ is shown in fig. \ref{fig:glauber}. It yields
$\sigma^{abs}_{\eta N}\approx$ 30 mb for the average value of 
$\alpha\approx 0.66$. In a similar analysis, recently Kotulla et al. 
\cite{Kotulla_08} have investigated the scaling behavior of the photoproduction 
of $\omega$-mesons off nuclei. They found typical absorption cross sections
in the range of 50 mb, corresponding to inelastic in-medium widths of the
mesons around 150 - 200 MeV. Here, we do not folluw-up this analysis
quantitatively, however, it is evident that also the extracted $\eta$ absorption
cross section must correspond to inelastic widths at least in the few ten MeV
range.
 
Due to the contribution from $\eta\pi$ final states and secondary production
processes, the inclusive reaction cannot be used to extract an in-medium
line shape of the S$_{11}$(1535) resonance.
A separation of quasi-free single $\eta$ production can only be achieved by 
cuts on the reaction kinematics.
\begin{figure}[thb]
\resizebox{0.50\textwidth}{!}{%
  \includegraphics{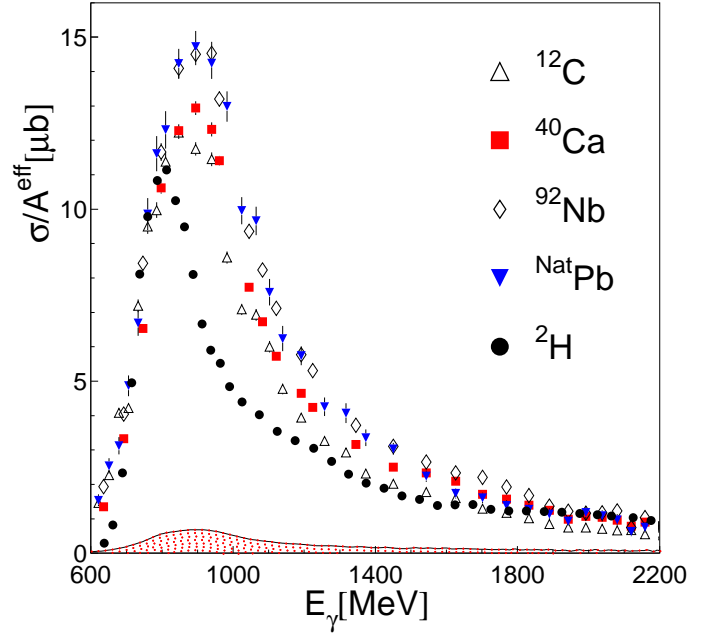}
}
  \caption{Total $\eta$ cross section with missing mass cut at 
140 MeV. Errors are statistical, the shaded band shows the systematic
uncertainty (excluding the 10\% flux normalization) for Calcium.}
\label{fig:single_eta}       
\end{figure}
A cut on the missing mass (see fig. \ref{fig:mismas}) at $\Delta m< 140$ MeV
is motivated by a comparison of the missing mass spectra to the simulated
line shape of the response for quasi-free single $\eta$ production.
The total cross sections after the cut are summarized in fig.
\ref{fig:single_eta}.

The shape (position and width) of the S$_{11}$-resonance structure is very 
similar for carbon, calcium, niobium, and lead.
A clear systematic evolution with mass number is not observed.
The shape is different for the deuteron data but this effect can mostly be
explained by the different momentum distributions of nucleons bound in the
deuteron or a heavy nucleus.

With the exception of the deuteron target, the separation of single quasi-free
$\eta$ production from other processes with the missing mass cut is only an
approximation, due to the overlapping tails of the distributions from
different processes. However, it gives already an indication,
that no strong in-medium effects on the shape of the S$_{11}$ resonance occur.
A more detailed discussion is possible by a comparison to the results of the
BUU-model subjected to the same kinematical cuts.

\section{Comparison to BUU results}
\label{chap:buucom}

\begin{figure}[th]
\resizebox{0.49\textwidth}{!}{%
\includegraphics{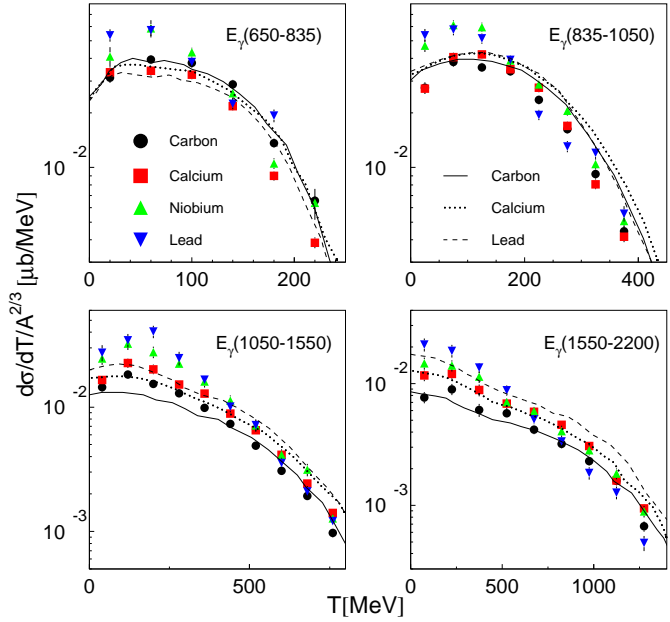}
}

\vspace*{0.5cm}
\resizebox{0.49\textwidth}{!}{%
\includegraphics{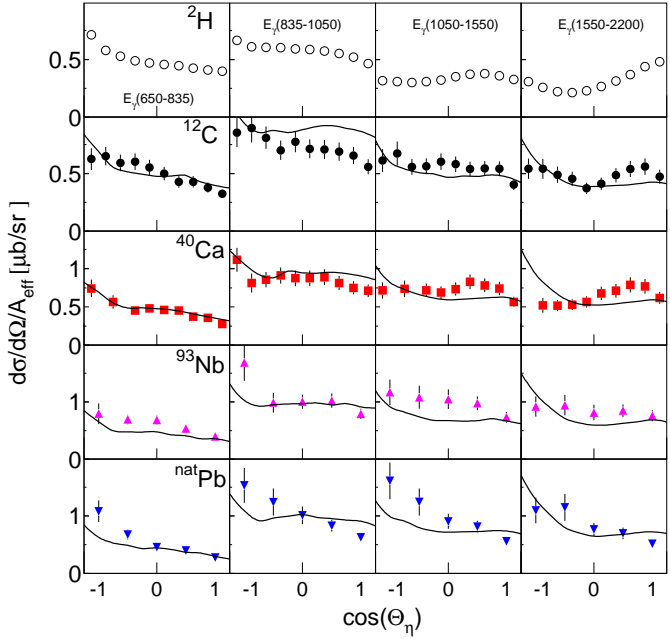}
}
\caption{Upper part: energy distributions for different incident photon beam
energy ranges. 
Bottom part: angular distributions. Curves are BUU model calculations.
Uncertainties include systematic effects except the 10\% flux normalization.
}
\label{fig:incl_diff}       
\end{figure}

\begin{figure}[tbh]
\resizebox{0.50\textwidth}{!}{%
  \includegraphics{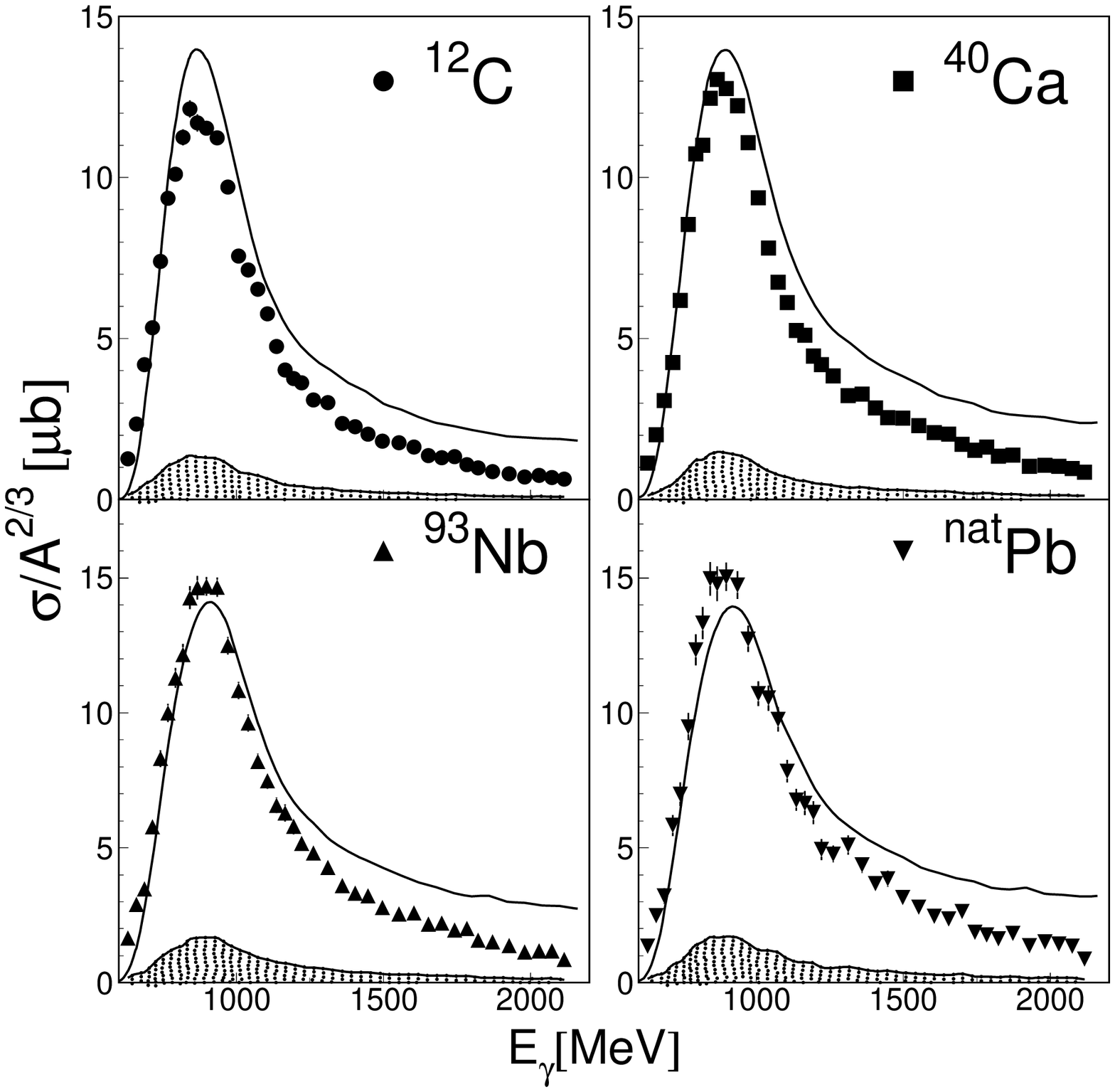}
}
\resizebox{0.50\textwidth}{!}{%
  \includegraphics{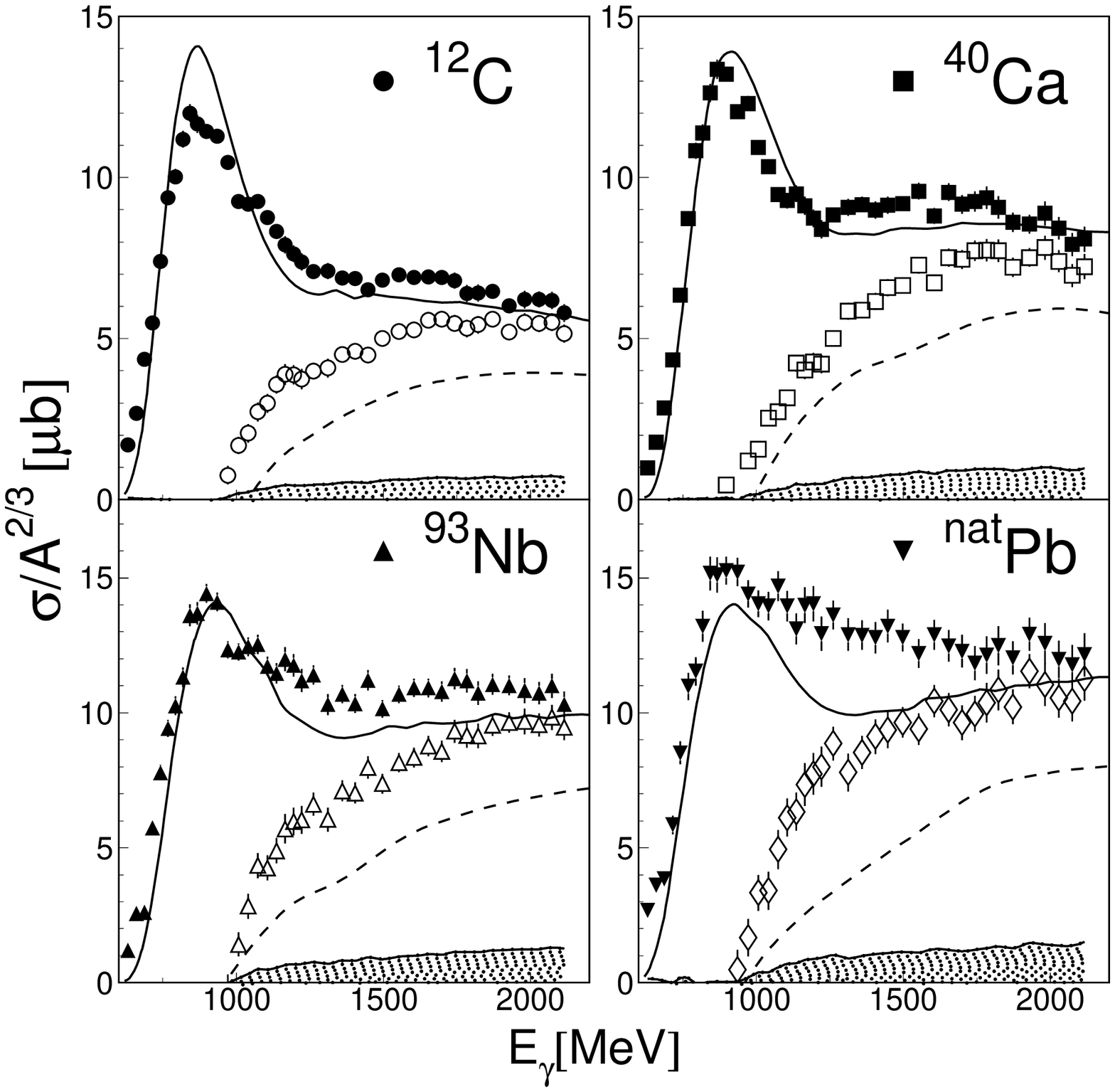}
}
\caption{Upper part: Comparison of the total exclusive single $\eta$
production cross (missing mass cut at 140 MeV) to BUU results.
Shaded bands: total systematic uncertainty.
Bottom part: filled symbols: inclusive cross sections (solid curves: BUU
results). Open symbols: difference of inclusive and exclusive single
$\eta$ production (dashed curves: BUU-results). Shaded bands: total systematic
uncertainty of the open symbols.}
\label{fig:mismas_1}       
\end{figure}

The distributions of kinetic energy and cm-angle (cm system of the incident
photon and a nucleon at rest) for the inclusive data are compared in Fig.
\ref{fig:incl_diff} to the results of the BUU model. The overall agreement
between data and model is quite good. The most significant disagreement is 
observed for the differential cross sections at small kinetic energies of the
$\eta$-mesons. In this regime, the model calculations are closer to the simple
$A^{2/3}$ scaling and underestimate the observed cross sections for the heavy
nuclei. As discussed below, these discrepancy can be traced back to the
contribution from $\eta\pi$ final states and/or secondary production processes
of $\eta$-mesons.
At low incident photon energies the angular distributions are similar to the
measured deuteron distributions (i.e. to the average nucleon coss section) 
and agree quite well with the model results. At the highest incident photon
energies, the angular distribution for the deuteron peaks at forward angles,
since for the free nucleon $t$-channel processes become important. This
effect is not seen for the heaviest nuclei, where the distributions peak at
backward angles. The model results show the same tendency and in the model the
backward peaking contribution arises mainly from secondary production processes.
This behavior is easily understood since $\eta$-mesons from secondary production
processes on average have small kinetic energies and therefore appear at
backward angles in the fast forward moving photon - nucleon cm-system. 
 
The total cross sections for the inclusive reaction, for quasi-free single 
$\eta$ production and for the contribution from $\eta\pi$ final states and
secondary production processes are summarized and compared to the model results
in Fig. \ref{fig:mismas_1}. The shape of the total inclusive cross section
is reasonably well reproduced for the lighter nuclei, but disagrees
significantly for lead, where the model shows still the peak of the S$_{11}$
resonance, which is absent in the data. However, this systematic shape change 
from light to heavy nuclei is not related to an in-medium modification of the 
S$_{11}$ resonance. This is clearly demonstrated by the separation of the
inclusive cross section into quasi-free single $\eta$-production and the other
components. The separation has been done by the missing mass cut at  
$\Delta m<$ 140 MeV, which has been applied to data and model calculations
(to the latter after folding them with the experimental resolution). Although, 
as discussed above, this separation is not perfect, the result for the 
line-shape of the S$_{11}$ dominating quasi-free single $\eta$ production is 
clear. Position, width, and peak cross section of the S$_{11}$  agree for 
all nuclei quite well with the model results. Only the peak cross section
shows a little systematic evolution from carbon (slightly overestimated)
to lead (slightly underestimated), which is however within the systematic
uncertainty introduced by the missing mass cut. The shape is fully explained 
by the `trivial' in-medium effects included in the BUU model, in particular 
the momentum distribution of the nucleons bound in heavy nuclei. 

The agreement is less good in the energy range above the S$_{11}$ resonance,
where the model overestimates the measured cross sections. However, one must 
keep in mind, that here the separation by the missing mass cut becomes strongly
dependent on the exact shape of the missing mass distributions for the different
components. Indeed the main effect at higher incident photon energies seems
to be a strong underestimation of the contribution from $\eta\pi$ final states
and/or secondary processes (see Fig. \ref{fig:mismas_1}, bottom part). 
In particular for lead around 1~GeV incident photon energy, this
component rises much more rapidly in the data than in the model.

\begin{figure}[tbh]
\resizebox{0.50\textwidth}{!}{%
  \includegraphics{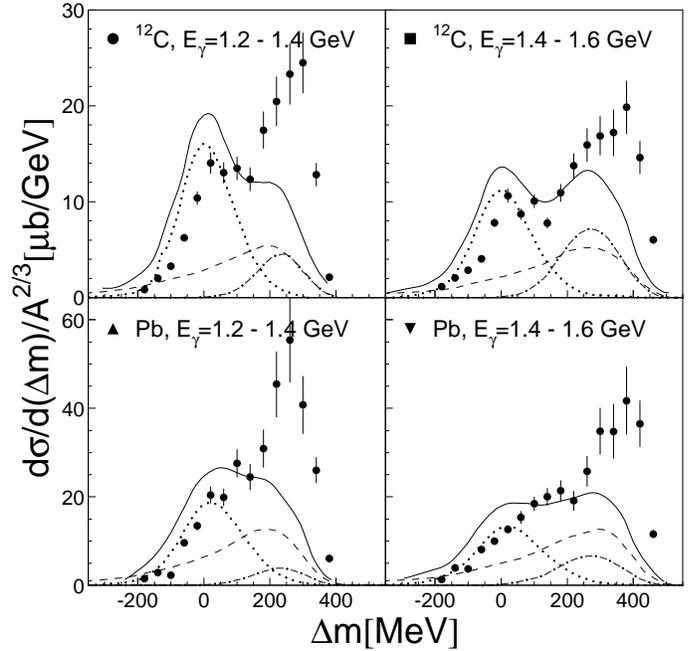}
}
\caption{Missing mass spectra for carbon and lead for two ranges of incident
photon energies. Curves: BUU-results for full model (full curves),
single, quasi-free $\eta$-production (dotted), $\eta\pi$ final states
(dash-dotted) and secondary $\eta$-production (dashed). }
\label{fig:cal_miss}       
\end{figure}

This mismatch is most clearly seen in the missing mass spectra in this
energy range which are compared in fig. \ref{fig:cal_miss} to the BUU
calculation. The contribution from single, quasi-free $\eta$ production
is overestimated while the contribution from $\eta\pi$ final
states seems to be underestimated in the model.
Part of this discrepancy is probably due to the uncertainty in the elementary
cross sections for $\eta\pi$ production reactions. There are recent 
precise data for the $\gamma p\rightarrow p\eta\pi^o$ reaction \cite{Horn_07},
however, much less is known for the channels with charged pions in the final
state or a neutron in the initial state.
In this range of incident photon energy, the modeled missing mass plots seem 
to indicate that the quasi-free single $\eta$-peak has already significant 
contamination from the tails of the `background' processes, in particular, 
from secondary $\eta$-production. However, it is not possible to obtain a
reasonable fit of the measured missing mass distributions by a variation
of the areas of the three model contributions, keeping their shape. Fitting 
the part of large missing mass with the $\eta\pi$ and secondary $\eta$ 
production contributions leads to unreasonable contributions of their tails
in the quasi-free region around zero missing mass. Therefore not only the
magnitude but also the shape of this contributions seems to be partly in 
conflict with the data. On the other hand, the missing mass shape of the 
quasi-free single $\eta$ production seems to be in better agreement with the
data (see Fig. \ref{fig:mismas}), it certainly agrees with it below the 
$\eta\pi$ threshold.  

\section{Conclusions}

The investigation of inclusive and exclusive $\eta$ production cross sections
for heavy nuclei from threshold to 2~GeV can be summarized as follows.
In the excitation region of the S$_{11}$(1535) resonance, contributions
from $\eta\pi$ final states and secondary production processes to inclusive
$\eta$ production are already significant.
At higher energies these contributions even become dominant.
A discussion of in-medium properties of the S$_{11}$ resonance or of absorption
properties of $\eta$ mesons in nuclear matter requires a careful treatment
of these effects.

An analysis of the scaling
of the cross sections with atomic mass number has been performed for $\eta$ mesons
produced closely to the kinematical limit where only quasi-free single $\eta$
production can contribute. Combined with previous low energy results
\cite{Roebig_96}, it is found that the scaling coefficient $\alpha$ is almost
constant at a value of 2/3 for $\eta$ kinetic energies from 20~MeV up to 1~GeV.
Using a simple Glauber model approximation, this corresponds to a constant
$\eta N$ absorption cross section of $\approx$30 mb.
A decrease of the absorption probability for $\eta$-mesons with kinetic energies
much above the S$_{11}$ range is not observed.

An analysis of the line shape of the S$_{11}$ resonance can be achieved with
the results for single, quasi-free $\eta$ production after cuts on the
reaction kinematics. The observed excitation functions for heavy nuclei
have almost identical shape from carbon to lead. The results
in the S$_{11}$ range are in good agreement with BUU model calculations
which include the `trivial' in-medium effects like Fermi smearing, Pauli
blocking of final states, and contributions from secondary processes.
Thus, an indication of a shift or broadening of the resonance has not been
found.

At higher incident photon energies, the agreement between BUU calculations
and experiment is less good. The relative contribution of single, undisturbed
$\eta$ photoproduction is overestimated in the model and the contribution
of secondary processes and/or $\eta\pi$ final states is significantly
underestimated. This indicates a need for better input for the semi-inclusive
$\eta X$ channels in the BUU calculations.

\section{Acknowledgments}
We wish to acknowledge the outstanding support of the accelerator group
and operators of ELSA.
This work was supported by Schweizerischer Nationalfonds and
Deutsche Forschungsgemeinschaft (SFB/TR-16.)

\end{document}